\documentclass[aps,twocolumn, amsmath,amssymb,pra,superscriptaddress]{revtex4-2}

\usepackage{graphicx}
\usepackage{dcolumn}
\usepackage[dvipsnames]{xcolor}
\usepackage{bm}

\usepackage{empheq,etoolbox}
\usepackage{bbold}

\usepackage[section]{placeins}

\newcommand{\p}{\partial}

\renewcommand{\i}{i}

\usepackage{verbatim}
\usepackage{xcolor}
\definecolor{lightgrey}{rgb}{0.8, 0.8, 0.8}
\definecolor{misprint}{rgb}{0, 0, 1}
\definecolor{remark}{rgb}{0.0, 0.5, 0.69}

\newcommand{\changes}[1]{\textcolor{black}{#1}}

\begin{document}

\title{Generalized Josephson effect in an asymmetric double-well potential\\ at finite temperatures}
\author{Kateryna Korshynska}
\affiliation{Department of Physics, Taras Shevchenko National University of Kyiv, 
64/13, Volodymyrska Street, Kyiv 01601, Ukraine}
\affiliation{Fundamentale Physik für Metrologie FPM, Physikalisch-Technische Bundesanstalt PTB, Bundesallee 100, 38116 Braunschweig, Germany}
\author{Sebastian Ulbricht}
\affiliation{Fundamentale Physik für Metrologie FPM, Physikalisch-Technische Bundesanstalt PTB, Bundesallee 100, 38116 Braunschweig, Germany}
\affiliation{Institut für Mathematische Physik, Technische Universität Braunschweig, Mendelssohnstraße 3, 38106 Braunschweig, Germany}

\begin{abstract} 
We investigate a non-interacting many-particle bosonic system, placed in an  asymmetric double-well potential. We first consider the dynamics of a single particle and determine its time-dependent probabilities to be in the left or the right well of the potential. These probabilities obey the standard Josephson equations, which in their many-particle interpretation also describe a globally coherent system, such as a Bose-Einstein condensate.
 This system exhibits the widely studied Josephson oscillations of the population imbalance between the wells. 
In our study we go beyond the regime of global coherence by developing a formalism based on an effective density matrix.
This formalism gives rise to a generalization of Josephson equations, which differ from the standard ones by an additional parameter, that has the meaning of the degree of fragmentation.
We first consider the solution of the generalized Josephson equations in the particular case of thermal equilibrium at finite temperatures, and extend our discussion to the non-equilibrium regime afterwards. 
Our model leads to a constraint on the maximum amplitude of Josephson oscillations for a given temperature and the total number of particles.
A detailed analysis of this constraint for typical experimental scenarios is given.
\end{abstract}

\maketitle

\section{Introduction}

In this work we study many-particle bosonic systems, as they are investigated in cold atom physics with the help of modern cooling and trapping techniques \cite{pethick2008bose}. 
At low temperatures bosons in a weakly interacting atomic gas can be regarded as wave packets. When approaching critical temperature, those wave packets begin to overlap until they form a single matter wave at $T = 0$ \cite{ketterle2002nobel}. 
This produces a unique state of matter known as Bose-Einstein condensate (BEC) \cite{pitaevskii2016bose}. Since the first experimental realization of a BEC \cite{anderson1995observation, davis1995bose}, its properties have been intensively studied in theory and experiment \cite{unnikrishnan2018bose, Smerzi1997, ketterle2002nobel, bloch2008many, bloch2005ultracold, gross1961structure, pitaevskii1961vortex, RevModPhys.71.S318} and led to various scientific applications \cite{ghosh2019sub, unnikrishnan2018bose, chiow2023ultra, ockeloen2013quantum, edwards2013atom, sewell2010atom, rudolph2015high}.

The quantum nature of a BEC manifests itself, when the gas is placed in a double-well potential.
Due to the coherence between the bosons in both wells, such a system exhibits a unique interference phenomenon, known as the Josephson effect \cite{RevModPhys.73.307,Smerzi1997, RevModPhys.71.463, Salasnich_1999, cordes2001tunnelling, PhysRevA.55.4318, hasegawa2013gaussian, song2008tunneling, Wang_2007, Bidasyuk, PhysRevA.90.043610}. This effect gives rise to collective oscillations of the particles, implying a time-dependent population imbalance between the wells. This behaviour has been widely studied \cite{RevModPhys.71.463, Salasnich_1999, cordes2001tunnelling, PhysRevA.55.4318, hasegawa2013gaussian, song2008tunneling, Wang_2007, Bidasyuk, PhysRevA.90.043610} and confirmed by the first realization of a single bosonic Josephson junction in 2005 \cite{Albiez_2005}.

Due to the numerous difficulties in realizing ultra-cold temperatures, nowadays BECs are accessible almost exclusively for laboratory experiments in fundamental physics \cite{unnikrishnan2018bose}. However, modern presicion metrology can also be performed with just laser cooled atoms, not being in BEC state, since coherence is not always an essential requirement in this area \cite{unnikrishnan2018bose}. This in particular makes cooled atom-based precision tools accessible for applications in technology. The aim of this work, therefore, is to study both the ultra-cold BEC and the finite temperature regime apart from the BEC state.

In the ultra-cold BEC regime, oscillations in a double-well are theoretically described by Josephson equations, which are derived under the idealized assumption of zero temperature $T = 0$.
Thus, in the present study we want to answer a natural question: how are these equations generalized for non-zero temperatures $T > 0$ and which additional effects do they predict?  

The paper is organized as follows. In Sec.~\ref{sec:model} we introduce the general features of a symmetric and an asymmetric double-well and define the relevant properties of its two lowest single-particle states. Based on this we derive the standard Josephson equations, which can be applied to describe a many-particle system in a globally coherent state.
Then in Sec.~\ref{sec:many paricle description} the single-particle states are used to build up the basis for the density matrix of a bosonic ensemble. 
From that in Sec.~\ref{subsec:effective density matrix description} we obtain 
an effective density matrix, that can be used to calculate the expectation values of 
one-particle operators.
With the help of this matrix we 
then derive a generalization of Josephson equations and analyze the additional physical effects they imply in Sec.~\ref{subsec: Generalized Josephson equations}. While the generalized Josephson equations give rise to various modifications of the standard Josephson effect, in Sec.~\ref{sec: Generalized Josephson effect} we focus on a system in thermal equilibrium and the occurrence of Josephson oscillations between the wells in the non-equilibrium regime. The conclusions are made in Sec.~\ref{sec: Conclusions}.
\vfill

\section{A single quantum particle in a double-well potential}
\label{sec:model}

In this paper we aim to consider a non-interacting many-particle bosonic system, which is a reasonable approximation of weakly interacting bosonic gases in many experimental situations \cite{fattori2008atom, Smerzi1997, bloch2008many, gross1961structure, pitaevskii1961vortex, bloch2005ultracold}.
This allows us to start our discussion at the level of single-particle states in this section and to construct from that a density matrix for the many-particle system afterwards.
\subsection{The symmetric double-well}
\label{subsec:single paricle states sym}

In what follows we want to keep our considerations as general as possible, making them applicable to a wide range of potentials that can be realized in experiment. 
For convenience we will restrict our analysis to one spatial dimension, 
\changes{assuming that the confinement in the other two directions is sufficiently strong, such that excitations of perpendicular modes are highly suppressed (see Appendix~\ref{Appendix: Thermalization in a bosonic gas}).} 

\begin{figure}[b]
    \centering
    \includegraphics[scale=0.4]{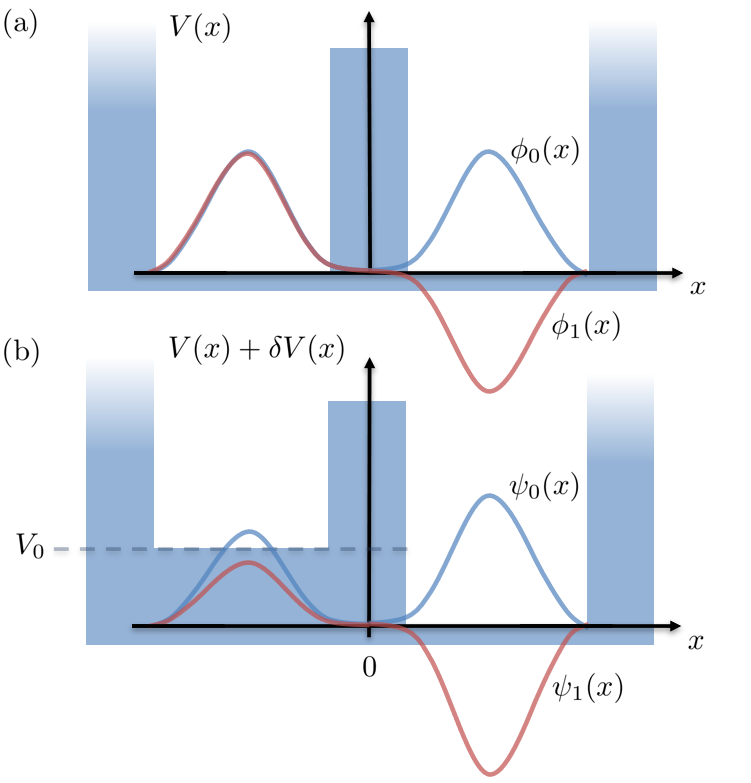}
   \caption{
  Example geometries of (a) a symmetric double-well potential $V(x)$ with the two lowest energy eigenstates $\phi_{0/1}(x)$ of a contained quantum particle, and (b) an asymmetric double-well $V(x)+\delta V(x)$ with corresponding states $\psi_{0/1}(x)$.}
   \label{fig: potential shape}
\end{figure}

We assume a symmetric, time-independent potential $V(x)=V(-x)$ without further specification of its explicit shape. This potential enters the single-particle Hamiltonian \begin{equation}
\hat H_0= -\frac{\hbar^2}{2m}\p_x^2 + V(x)\,. \label{eqn:one_particle_Hamiltonian}
\end{equation}
The corresponding eigenvalue problem $\hat H_0 \phi(x)=E\phi(x)$ can be solved in terms of orthogonal basis functions $\phi_n(x)$ with energies $E_n$. We assume that the Hamiltonian has at least two normalized eigenstates: the ground state $\phi_0(x)$ and the first excited state $\phi_1(x)$. Since the potential $V(x)$ is symmetric, the ground state is necessarily symmetric as well, while the first excited state must be anti-symmetric, as visualized in Fig.~\ref{fig: potential shape}a. 
From these states we want to construct proper left and right states $\phi_{L/R}(x)$, which describe the quantum particle being in the left or right part of the potential. To do this we linearly combine the energy eigenstates in the form 
\begin{eqnarray}
  \phi_L(x)&=&\cos\xi \,\phi_0(x)+\sin\xi \,\phi_1 (x) \label{eqn:LP_states} \\
  \phi_R(x)&=&\sin\xi \,\phi_0(x)-\cos\xi \,\phi_1 (x)\,\nonumber
\end{eqnarray}
with a parameter $\xi$ that needs to be determined hereafter.
To investigate these states, we now introduce the left- and right-side scalar products
\begin{equation}
\langle \,\cdot\,|\,\cdot\,\rangle_L+\langle \,\cdot\,|\,\cdot\,\rangle_R=\langle \,\cdot\,|\,\cdot\,\rangle\,,
\end{equation}
which imply an integral over all negative or positive $x$, respectively, such that their sum gives the standard scalar product.
Calculating the left- and right-side scalar products of the left and right states (\ref{eqn:LP_states}) we obtain
\begin{eqnarray}
    \langle \phi_R|\phi_R\rangle_R =\langle \phi_L|\phi_L\rangle_L &=& 1/2+\sin(2\xi)\langle \phi_0|\phi_1\rangle_L\nonumber \\
    \langle \phi_R|\phi_R\rangle_L = \langle \phi_L|\phi_L\rangle_R&=& 1/2-\sin(2\xi)\langle \phi_0|\phi_1\rangle_L\,. \label{eqn:LRscalar_products}
\end{eqnarray}
In this calculation we use $\langle \phi_0|\phi_0\rangle_{L/R} =\langle \phi_1|\phi_1\rangle_{L/R} =1/2$, which is clear by symmetry considerations.
The scalar product $\langle \phi_0|\phi_1\rangle_L=-\langle \phi_0|\phi_1\rangle_R$
can be chosen real and positive by adapting the phases of the energy eigenstates. 
Recalling that the scalar products (\ref{eqn:LRscalar_products}) represent the probabilities to find the particles of state $\phi_{L/R}(x)$ in the left or right part of the potential, we can maximize this probability by the choice  $\xi=\pi/4$, leading to the states
\begin{eqnarray}
\phi_L(x)&=&\frac{1}{\sqrt{2}}[\phi_0(x)+\phi_1 (x)]\nonumber
\\
\phi_R(x)&=&\frac{1}{\sqrt{2}}[\phi_0(x)-\phi_1 (x)]\, \label{eqn:optimal_states}
\end{eqnarray}
as the best fitting left and right states in the potential $V(x)$, shown in Fig.~\ref{fig: left and right states}a. The explicit shape of this potential now only enters our consideration via the scalar product $\langle \phi_0|\phi_1\rangle_L$, which is a measure of how good the states $\phi_{L/R}(x)$ actually represent a particle at the left or right side of the potential. In particular, we will call all such potentials a symmetric \emph{double-well}, in which  $\langle \phi_0|\phi_1\rangle_L=1/2-\epsilon$ with a sufficiently small $\epsilon>0$, such that
\begin{eqnarray}
    \langle \phi_R|\phi_R\rangle_R =\langle \phi_L|\phi_L\rangle_L &=& 1-\epsilon\nonumber\\
    \langle \phi_R|\phi_R\rangle_L = \langle \phi_L|\phi_L\rangle_R&=&\epsilon \,,\label{eqn:well_defined}
\end{eqnarray}
and the $\phi_{L/R}(x)$ appropriately describe a single quantum particle in a left or right well state.

In the next section, we will use these definitions for the symmetric double-well to extend our analysis to a double-well potential with an additional potential step.

\subsection{The asymmetric double-well}

\label{subsec:single paricle states asym}
To describe a particle in an asymmetric double-well potential we extend the Hamiltonian 
\begin{equation}
    \hat H = \hat H_0 + \delta V(x) \label{eqn:full_single_particle_Hamiltonian}
\end{equation} by a small potential step
\begin{equation}
\delta V(x)= \left\{\begin{array}{ll}
        V_0& \quad x<0 \\
       0  & \quad \mathrm{elsewhere}\,\,.
       \end{array}\right.\,
       \label{Eq: perturb V0}
\end{equation}
The solutions of the corresponding eigenvalue problem
\begin{equation}
\hat H \psi_n(x)=\tilde E_n \psi_n(x) \label{eqn:perturb_eigenvalue_problem}
\end{equation}
can be obtained in the framework of first order perturbation theory, up to linear order in $V_0/E$, where $E=(E_0+E_1)/2$. For that we use the states from Sec.~\ref{subsec:single paricle states sym} and treat $\delta V(x)$ as a small perturbation of the symmetric double-well. 

\begin{figure}[b!]
    \centering
\includegraphics[scale=0.35]{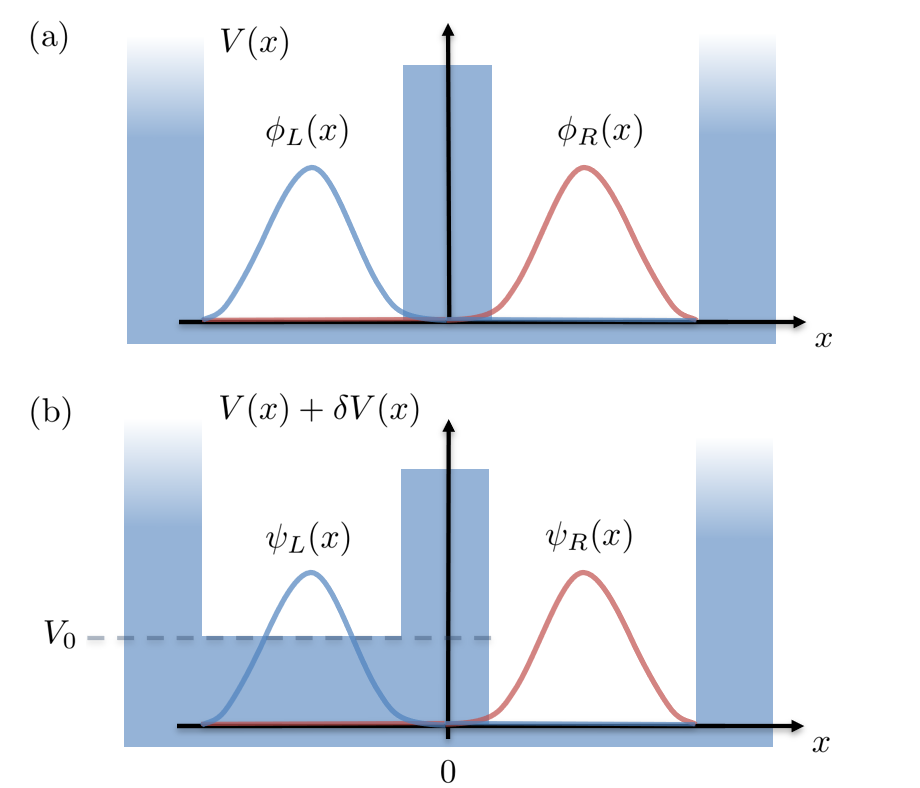}
   \caption{Left and right well states in (a) a symmetric  and (b) an asymmetric double-well potential. The functional form of both coincides, regardless of the value of $V_0\ll E$. }
   \label{fig: left and right states}
\end{figure}

It is well known that in perturbation theory, states strongly affect each other when they have similar energies. 
For the symmetric double-well (\ref{eqn:well_defined}) 
the separation of the ground and excited state from other possible states is in the order of $E$. In contrast, their own energies are much closer to each other, i.e., $E_1-E_0=\Delta E\ll E$. 
This is due to the fact that the scalar product $\langle \phi_0|\phi_1\rangle_L=1/2-\epsilon$ only slightly deviates from $1/2$, such that the states $\phi_0(x)\approx\phi_1(x)$ closely resemble each other for $x<0$, as also can be seen in Fig.~\ref{fig: potential shape}a.
In consequence, acting on them with the Hamiltonian $\hat H_1$ gives similar energy eigenvalues $E_0\approx E_1$, which become degenerated in the ultimate case $\epsilon=0$.
We, thus, can restrict our investigation to the corrections of $\phi_0(x)$ by $\phi_1(x)$ and vise versa, while we have to deal with the perturbation theory for nearly degenerated states \cite{binney2013physics}.
For that we make the ansatz ${\psi}_n(x) = c^0_n\phi_0(x) + c^1_n\phi_1(x)$ in Eq.~(\ref{eqn:perturb_eigenvalue_problem}) and integrate this equation with the states $\phi^*_{0}(x)$ and $\phi^*_{1}(x)$, respectively. This, together with the normalization of the corrected states, yields six equations for the six constants $c^0_n$, $c^1_n$, and $\tilde E_n$ with $n=0,1$. The only non-trivial, linearly independent solution is given by
\begin{eqnarray}
   \psi_{0}(x) &=& \mathcal{C}\left( \phi_{0}(x) - \frac{V_0}{\Delta E+\sqrt{\Delta E^2 + V_0^2}}\,\phi_1(x) \right)\label{eqn:asymm_energy_states}\\
      \psi_{1}(x) &=& \mathcal{C}\left( \phi_{1}(x) + \frac{V_0}{\Delta E+\sqrt{\Delta E^2 + V_0^2}}\,\phi_0(x) \right) \nonumber \,
\end{eqnarray}
with \mbox{$\mathcal{C}=\frac{1}{\sqrt{2}}\left(1+\Delta E/\sqrt{\Delta E^2 + V_0^2}\right)^{1/2}$}.
Those states satisfy Eq.~(\ref{eqn:perturb_eigenvalue_problem}) to linear order in $V_0/E$ for the corrected energies
\begin{equation}
        \tilde{E}_{0/1} =E+\frac{V_0}{2}\mp\frac{1}{2}\sqrt{\Delta E^2+V_0^2}\,,
        \label{eq: tilde E0/1}
\end{equation}
where we assume that $\epsilon V_0/E$ can be neglected, cf. Eq.~(\ref{eqn:well_defined}).
The $\psi_{0/1}(x)$, hence, are energy eigenstates of the asymmetric double-well,
illustrated in Fig.~\ref{fig: potential shape}b. 

As for the symmetric double-well in the last section, we now can define the left and right well states $ \psi_{L/R}(x)$ in the case of the asymmetric double-well potential
\begin{eqnarray}
   \psi_L(x)&=&\cos\xi \, \psi_0(x)+\sin\xi \, \psi_1 (x)  \nonumber\\
  \psi_R(x)&=&\sin\xi \, \psi_0(x)-\cos\xi \,\psi_1 (x)\,\label{eqn:LP_states_asymm}
\end{eqnarray}
in total analogy to Eq.~(\ref{eqn:LP_states}), but
this time using the energy eigenstates $\psi_{0/1} (x)$ as a basis.

We now again ask for the optimal states that maximize $\langle \psi_R| \psi_R\rangle_R$ and $\langle \psi_L| \psi_L\rangle_L$. Inserting Eq.~(\ref{eqn:asymm_energy_states}) into Eq.~(\ref{eqn:LP_states_asymm}) we can formulate this question in the basis $\phi_{0/1} (x)$ of the symmetric double-well problem.
In this basis the optimal states $\phi_{L/R}(x)$ are already known and given by Eq.~(\ref{eqn:optimal_states}).
Therefore, we find that those states are optimal for both, the symmetric and the asymmetric case, i.e,
$\psi_L(x)=\phi_L(x)$ and $ \psi_R(x)=\phi_R(x)$, see Fig.~\ref{fig: left and right states}.
In Eq.~(\ref{eqn:LP_states_asymm}) this is achieved by the choice
\begin{equation}
    \xi =\arcsin\left[\frac{1}{\sqrt{2}}\left(1+V_0/\sqrt{\Delta E^2 + V_0^2}\right)^{1/2}\right]\,, \label{eqn:xi_of_V0}
\end{equation} which for $V_0=0$ recovers the symmetric case $\xi=\pi/4$.

Having found the optimal left and right states for the asymmetric double-well potential, we can investigate their dynamics, which is governed by the Hamiltonian operator (\ref{eqn:full_single_particle_Hamiltonian}). With $\psi_{L/R}(x)$ being a superposition of energy eigenstates, it is clear that they do not satisfy an eigenvalue equation for a particular energy. Instead, we obtain the coupled equations
\begin{eqnarray}
\hat H  \psi_{L}(x)  &=& E_{L} \psi_{L}(x)  + K  \psi_{R} (x) \nonumber \\
\hat H  \psi_{R}(x) &=&  E_{R} \psi_{R}(x)+  K  \psi_{L} (x)\label{eq: psiLR single particle}\,,
\end{eqnarray}
with the newly introduced constants 
\begin{equation}
E_{L}=E+V_0\, ,\quad 
E_R=E\, ,\quad K=-\frac{\Delta E}{2}\,. \label{eqn: constants}
\end{equation}
With that we find a coupling $\sim \Delta E$ between the left and right state, which vanishes when the energies $E_{0/1}$ degenerate, as it is the case when the wells are seperated by a very high potential barrier.
Moreover, the difference 
$E_{L} - E_{R} =V_0$ coincides with the potential step (\ref{Eq: perturb V0}), which is a well known feature of a Josephson junction \cite{dauphinee2015asymmetric}.

\subsection{Josephson equations and many-particle interpretation}
\label{subsec: Josephson equations and many-particle interpretation}
In the last section we analyzed the eigenvalue equation $\hat H\psi_{0/1}= \tilde{E}_{0/1}\psi_{0/1}$, holding for the time-dependent, but stationary states $\psi_{0/1}(x,t)=\exp(-i \tilde{E}_{0/1} t/\hbar)\psi_{0/1}(x)$.
From the spatial wave functions $\psi_0(x)$ and $\psi_1(x)$ the left and right states $\psi_L(x)$ and $\psi_R(x)$ were constructed, cf. Eqs.~(\ref{eqn:LP_states_asymm}).
Now we want to investigate the evolution of the states $\psi_L(x,t)$ and $\psi_R(x,t)$, which are not stationary states of a particular energy $E_{L/R}$, solely, as becomes clear in Eqs.~(\ref{eq: psiLR single particle}).
Instead, we can assume a time-dependent superposition of $\psi_L(x)$ and $\psi_R(x)$, giving rise to a single wave-function
\begin{equation}
   \psi(x,t) = w_L(t) \psi_{L}(x) + w_R(t) \psi_{R}(x) \,. \label{eqn:pure_state_wave_function}
\end{equation}
By using the action  (\ref{eq: psiLR single particle}) of the one-particle Hamiltonian $\hat{H}$ on the left and right well states, as well as the orthogonality relation $\langle \psi_{L}| \psi_{R} \rangle = 0$, we obtain
\begin{eqnarray*}
   i\hbar \dot{w}_{L}(t) &=& E_{L}w_{L}(t) +  K w_{R}(t)\\
   i\hbar \dot{w}_{R}(t) &=& E_{R}w_{R}(t) +  K w_{L}(t).
\end{eqnarray*}
These equations describe the evolution of the coefficients $w_{L/R}(t)$, where
   $|w_{L/R}(t)|^2$ represent the time-dependent probabilities to find the particle in the left or right well, respectively.
Here, the normalization condition would be given by $|w_L(t)|^2 + |w_R(t)|^2 = 1$.

While, so far we have applied the wave function (\ref{eqn:pure_state_wave_function}) to describe a single particle, it can also be used to represent a many-particle system, as long as that system behaves coherently, i.e, is in a \emph{pure state}. 
In this case, we only need to renormalize Eq.~(\ref{eqn:pure_state_wave_function}) to obtain $|w_L(t)|^2 + |w_R(t)|^2 = N$, where $N$ is the total number of bosons.
Then, the $|w_{L/R}(t)|^2=N_{L/R}(t)$ represent the \emph{population} of the left and right well, respectively.

Staying with the many-particle interpretation and by defining $w_{L/R}(t) = \sqrt{N_{L/R}(t)}e^{i \theta_{L/R}(t)}$ we find the coupled differential equations
\begin{eqnarray}
   \hbar \dot{Z} &=& 2K \sqrt{1-Z^{2}}\sin \theta 
   \label{Eq: pure state Josephson1} \\
   \hbar \dot{\theta} &=& E_{L} - E_{R} - \frac{2KZ \cos \theta}{\sqrt{1-Z^{2}}},
   \label{Eq: pure state Josephson2}
\end{eqnarray}
for the fractional population imbalance $Z(t) = (N_{L}(t) - N_{R}(t))/N$ and the phase difference $\theta(t) = \theta_{R}(t) - \theta_{L}(t)$.
These are standard Josephson equations, as they are used in literature 
 to describe the standard Josephson effect \cite{RevModPhys.71.463, Salasnich_1999, cordes2001tunnelling, PhysRevA.55.4318, hasegawa2013gaussian, song2008tunneling, Wang_2007}. 
As we recapitulate in Appendix~\ref{Appendix: Standard Josephson equations}, the general solution of these equations leads to oscillations of the observable population imbalance $Z(t)$ with the frequency $\sqrt{\Delta E^2 + V_0^2}/\hbar$.

Having found these familiar expressions for a many-particle system in a pure state, in what follows we will investigate what happens when the pure state assumption is not initially made. 
We will show, that this leads to a generalization of Josephson equations, opening up a way to investigate many-particle dynamics in the double-well beyond the regime of global coherence.

\section{Many-particle bosonic system in a double-well}
\label{sec:many paricle description}

In the last section we have studied the dynamics of a single particle based on the generic features of a double-well potential giving rise to specific eigenstates and energies. The following ideas, however, can be applied to an arbitrary two-state system.

Having defined single-particle states $|\psi_{0/1}\rangle$ in a given geometry, we can elaborate a description of a many-particle system. For that we assume the Hamiltonian of a bosonic ensemble of $N$ non-interacting particles, given by
\begin{equation}
\hat{\mathcal{H}} = \sum_{i=1}^{N}\hat{H}(x_{i})\,
\label{eq: many-particle hamiltonian H}
\end{equation}
with the single-particle Hamiltonian (\ref{eqn:full_single_particle_Hamiltonian}). 
\changes{The assumption of non-interacting particles is valid, as long as the one-dimensional particle density is much smaller than the inverse of the scattering length for the considered bosons \cite{mazets2010thermalization}. The consequences of this assumption for the thermalization process are discussed in Appendix~\ref{Appendix: Thermalization in a bosonic gas}.}
In general, one would describe such a system by introducing the hermitian density operator \cite{landau2013statistical}
\begin{equation}
    \hat{\rho} =  \sum_{N_{1} = 0}^{N}\sum_{\tilde{N}_{1} = 0}^{N} p_{N_{1} \tilde{N}_{1}}(t) |\Psi_{N_{1}}\rangle \langle \Psi_{\tilde{N}_{1}}|\,,
    \label{Eq: density matrix}
\end{equation}
where $p_{N_{1} N_{1}}$ is the probability that the system is in a state with $N_{1}$ excited particles. The normalization condition for the density matrix reads $\sum_{N_{1} = 0}^{N}p_{N_{1} N_{1}} = 1$. Moreover, the state $| \Psi_{N_{1}} \rangle$ in Eq.~(\ref{Eq: density matrix}) is given by
\begin{multline}
|\Psi_{N_{1}}\rangle = \frac{1}{\sqrt{N!(N-N_{1})!N_{1}!}} \int dx_{1}...dx_{N} \\ \times \sum_{j(N_{1})}  \psi_{\sigma_{j}^{N_{1}}(1)}(x_{1})\psi_{\sigma_{j}^{N_{1}}(2)}(x_{2})...\psi_{\sigma_{j}^{N_{1}}(N)}(x_{N}) \changes{|x_{1},\dots,x_{N}\rangle}, 
\label{whole_state}
\end{multline}
where the sum is over all possible configurations $j(N_{1})$ of single-particle states for a fixed number $N_{1}$. 
Each configuration is labeled by an index $j$ and characterised by a vector $\mathbf{\sigma}_{j} = (\sigma_{j}(1), \sigma_{j}(2), ..., \sigma_{j}(N))$, defining the state of each particle. 

Since the state $|\Psi_{N_{1}}\rangle$ is constructed from the single-particle states $\psi_{0}(x)$  and $\psi_{1}(x)$, the $\sigma_{j}(i)$ can only take the values $0$ or $1$. However, the particle $i$ is still allowed to be in a superposition of both states, since the coefficients $p_{N_{1}, \tilde{N}_{1}}$ for $N_{1} \neq \tilde{N}_{1}$ are not necessarily zero.

\subsection{Effective density matrix description}
\label{subsec:effective density matrix description}

In this section we will reformulate the description of a many-particle system by the density matrix (\ref{Eq: density matrix}) in terms of an effective density matrix in the basis of the one-particle states $|\psi_{0}\rangle$ and $|\psi_{1}\rangle$, which appear as the two lowest states in the double-well potential. In the following we are particularly interested in the occupation probabilities of these states, needed to investigate the Josephson effect. 

For this purpose, we consider a generic single-particle operator 
\begin{multline*}
    \hat{\mathcal{O}} = \underbrace{\hat{O} \oplus \mathbb{1} \oplus \dots \oplus \mathbb{1}}_{N\textnormal{-times}} + \mathbb{1} \oplus \hat{O} \oplus \dots \oplus \mathbb{1} + \dots \\[-0.3 cm]+ \mathbb{1} \oplus \mathbb{1} \oplus \dots \oplus \hat{O}
\end{multline*}
in the $N$-particle Hilbert space, constructed from the operator $\hat{O}$, acting on a single-particle state. For instance, the occupation number operator $\hat{N}_{0/1}$ is given by the particular choice  $\hat{O} = |\psi_{0/1}\rangle \langle \psi_{0/1}|$ in the formula above. The expectation value of a many-particle operator then reads 
\begin{multline}
    \langle \hat{\mathcal{O}}\rangle_{\rho}  = \mathrm{tr}(\hat{\rho}\hat{\mathcal{O}}) = \sum_{N_{1} = 0}^{N}\sum_{\tilde{N}_{1} = 0}^{N}p_{N_{1}\tilde{N}_{1}} \langle \psi_{\tilde{N}_{1}}|\hat{\mathcal{O}} |\psi_{N_{1}}\rangle  \\ = \sum_{N_{1} = 0}^{N} p_{N_{1}N_{1}} [N_{1}\langle \psi_{1}|\hat{O} |\psi_{1} \rangle + (N-N_{1})\langle \psi_{0}|\hat{O} |\psi_{0}\rangle]  \\ + \sum_{N_{1} = 0}^{N-1}\sqrt{(N_{1} + 1)(N - N_{1})}[p_{N_{1} + 1, N_{1}}\langle \psi_{0}|\hat{O} |\psi_{1} \rangle  \\ + p_{N_{1}, N_{1} + 1}\langle \psi_{1}|\hat{O} |\psi_{0}\rangle ],
    \label{effective density recipe}
\end{multline}
cf. \cite{Gustafson2003}.
In front of the expectation values $\langle \psi_{i}| \hat{O} | \psi_{j} \rangle$ we now can read off the coefficients $\alpha_{ij}$, which contain all information accessible by the measurement of one-particle observables: 
\begin{eqnarray*}
    \alpha_{00} &=& \sum_{N_{1} = 0}^{N}(N- N_{1}) p_{N_{1}N_{1}}\\
    \alpha_{11} &=& \sum_{N_{1} = 0}^{N} N_{1} p_{N_{1}N_{1}}\\
    \alpha_{01}  &=& \sum_{N_{1} = 0}^{N-1} \sqrt{(N_{1} + 1)(N - N_{1})} p_{N_{1}+1, N_{1}}\\
     \alpha_{10} &=& \alpha^{*}_{01}.
\end{eqnarray*}

We now formally rearrange Eq.~(\ref{effective density recipe}) and introduce the effective density matrix $\hat{\rho}_\mathrm{e}$:
\begin{equation}
   \langle \hat{\mathcal{O}} \rangle_{\rho} = 
   \begin{bmatrix} 
	\langle \psi_{0} | & \langle \psi_{1} |  \\
	\end{bmatrix}    \begin{bmatrix} 
	\alpha_{00} & 	\alpha_{01}  \\
        \alpha_{10} & 	\alpha_{11}  \\ 
	\end{bmatrix} \hat{O} \begin{bmatrix} 
	|\psi_{0} \rangle  \\
        | \psi_{1} \rangle \\ 
	\end{bmatrix}  = \mathrm{tr}(\hat{\rho}_\mathrm{e}\hat{O})\,.
 \label{Eq: rho e definition}
\end{equation}
As we show in Appendix~\ref{Appendix: Interpretation of effective density matrix}, this effective density matrix, which we obtained from the calculation of one-particle expectation values, coincides with the reduced density matrix of a single particle in a bath of all the other $N - 1$ bosons \cite{erdahl2012density}. It therefore can be interpreted as a description of the many-particle ensemble by an average boson. 

Further, we are interested in the population of the left and right potential wells rather than the occupation of the global ground and excited states $|\psi_{0/1}\rangle$. Therefore, in the following we will describe the system in the basis of left and right well states $|\psi_{L/R} \rangle$, introduced in Sec.~\ref{sec:model}. 
Here this is done by the matrix transformation
\begin{equation}
        \begin{bmatrix} 
	|\psi_{L} \rangle  \\
        |\psi_{R} \rangle \\ 
	\end{bmatrix} =
       \begin{bmatrix} 
	\cos \xi |\psi_{0}\rangle + \sin \xi |\psi_{1} \rangle  \\
    \sin \xi   | \psi_{0} \rangle - \cos \xi |\psi_{1} \rangle \\ 
	\end{bmatrix} = \hat{T}\begin{bmatrix} 
	|\psi_{0}\rangle   \\
        | \psi_{1} \rangle \\ 
	\end{bmatrix}
 \label{eq:left-right states}\,,
\end{equation}
where the parameter $\xi$ is given by Eq.~(\ref{eqn:xi_of_V0}) for the asymmetric double-well potential.
The matrix $\hat{T}$ 
now can be used to express the effective density matrix $\hat{\rho}_{e LR} = \hat{T}\hat{\rho}_{e}\hat{T}^{-1}$ in the left and right well basis.
In general, this hermitian $2\times 2$ matrix can be parameterized by 
\begin{equation}
   \hat{\rho}_{\mathrm{e} LR} = \begin{bmatrix} 
	N_{L}(t) & 	A(t)e^{i\theta(t)}  \\
        A(t)e^{-i\theta(t)} & 	N_{R}(t)  \\ 
	\end{bmatrix},
 \label{Eq: effective density matrix}
\end{equation}
where $N_{L}(t)$ and $N_{R}(t)$ are the occupation numbers of the left and right well, respectively. Moreover, by $\mathrm{tr} (\hat{\rho}_{\mathrm{e} LR}) = N = N_{L}(t) + N_{R}(t)$ the total number of particles is conserved. The non-diagonal complex matrix elements describe the interference between left and right well states and therefore induce the coupling between the wells by a mixing parameter $A(t)$ and the phase difference $\theta(t)$ between them. 

In what follows, we will derive the equations of motion for $\hat{\rho}_{\mathrm{e} LR}$, which will generalize the standard Josephson equations (\ref{Eq: pure state Josephson1}) and (\ref{Eq: pure state Josephson2}) to the case where the system is not described by a single wave-function.

\subsection{Generalized Josephson equations}
\label{subsec: Generalized Josephson equations}

With the help of the effective density matrix (\ref{Eq: effective density matrix}), we can go beyond the description of a many-particle system by a pure state. This will give us the opportunity to consider a wider range of physical setups. For instance, this will enable us to describe bosonic systems at finite temperatures. 

The evolution of the effective density matrix is given by the Liouville equation
\begin{equation}
   i\hbar \frac{\partial}{\partial t}\hat \rho_{\mathrm{e} LR} = [\hat{\mathcal{H}}_{\mathrm{e} LR}, \hat{\rho}_{\mathrm{e} LR}],
   \label{Eq: Liouville equation}
\end{equation}
with the Hamiltonian operator $\hat{\mathcal{H}}_{\mathrm{e} LR} = \hat{H}|\psi_{L}\rangle \langle \psi_{L}| +\hat{H}|\psi_{R}\rangle \langle \psi_{R}|$, where the single-particle Hamiltonian $\hat H$ is given by Eq.~(\ref{eqn:full_single_particle_Hamiltonian}).
By using the action of $\hat{H}$ on the left and right well states $|\psi_{L/R}\rangle $ given in Eq.~(\ref{eq: psiLR single particle}), the Liouville equation reduces to the three coupled differential equations
\begin{eqnarray}
    \hbar \dot{Z} &=& 4\frac{K}{N} A \sin \theta
    \label{Eq: generalized Josephson1}\\
    \hbar \dot{\theta} &=& E_{L} - E_{R} - \frac{KNZ}{A}\cos \theta
    \label{Eq: generalized Josephson2}\\
    \frac{\dot{A}}{A} &=& \left[\dot{\theta} + \frac{E_{R} - E_{L}}{\hbar}\right] \tan \theta\,.
    \label{Eq: generalized Josephson3}
\end{eqnarray}

In what follows we want to draw attention to the additional physical effects that are described by the generalized Josephson equations (\ref{Eq: generalized Josephson1}) - (\ref{Eq: generalized Josephson3}) in comparison to the standard ones (\ref{Eq: pure state Josephson1}),(\ref{Eq: pure state Josephson2}). For this purpose, we use the fact that these equations can be rewritten in the form  
\begin{eqnarray}
   \hbar \dot{Z} &=& 2K \sqrt{f^{2}-Z^{2}}\sin \theta \label{eqs: generalized with f1}  \\
   \hbar \dot{\theta} &=& E_{L} - E_{R} - \frac{2KZ \cos \theta}{\sqrt{f^{2}-Z^{2}}},
   \label{eqs: generalized with f2}
\end{eqnarray}
as we show in Appendix~\ref{Appendix: Degree of fragmentation}. The structure of these equations is obtained by formally integrating Eq.~(\ref{Eq: generalized Josephson3}) for $A(t)$ and eliminating the mixing parameter in the Josephson equations (\ref{Eq: generalized Josephson1}) and (\ref{Eq: generalized Josephson2}) for $Z(t)$ and $\theta(t)$. The representation (\ref{eqs: generalized with f1}),(\ref{eqs: generalized with f2}) of the generalized Josephson equations only deviates from the standard ones by the newly introduced constant parameter $f$. At the level of the effective density matrix, $f$ is fixed by the initial condition $\hat{\rho}_{\mathrm{e}LR}(t=0)$.

For $f = 1$ we recover the standard Josephson equations, i.e., dynamics in the pure state regime. At the level of the effective density matrix this corresponds to the special choice $A(t) = \sqrt{N_{L}(t)N_{R}(t)} = N\sqrt{1- Z(t)^{2}}/2$. 
In this case the effective density matrix is a projector $|\psi\rangle\langle\psi|$ 
to one of its eigenvectors, representing a globally coherent state, such that 
$\langle \hat{\mathcal{O}}\rangle_{\rho}=\langle \psi| \hat{\mathcal{O}}|\psi\rangle$.
In this regime the system can be described by the single wave function $\psi$, given by Eq.~(\ref{eqn:pure_state_wave_function}).

For general values of $f$ 
the eigenvalues of the effective density matrix $\hat{\rho}_{\mathrm{e}}$ are given by $N(1 \pm f)/2$, see Appendix~\ref{Appendix: Degree of fragmentation}.
Recalling the interpretation of the effective density matrix as the reduced density matrix for an average boson, these eigenvalues have to be non-negative, to ensure non-negative probabilities. This leads to the restriction $|f| \leq 1$. 
Following the discussion of BEC fragmentation, given in \cite{PhysRevA.74.033612, PhysRevA.59.3868}, we choose $f>0$ and use the term \emph{degree of fragmentation} for the $1 - f$ parameter, implying $f=1$ for a non-fragmented globally coherent state and $f=0$ for two incoherent fragmented states.
\changes{
The $f$ parameter can be associated with the commonly used degree of coherence, or coherence factor, which defines the visibility of interference fringes in an interference experiment with fragmented BECs \cite{pitaevskii2016bose}. 
However, notice that the same terms are used differently in the studies of quantum fluctuations in BECs  \cite{pitaevskii2001thermal, PhysRevLett.96.130404}, where a strongly interacting system is discussed (see also Appendix~\ref{Appendix: Degree of fragmentation}).
}

\section{Generalized Josephson effect }
\label{sec: Generalized Josephson effect}

Let us now briefly summarize what we have done so far. We started with a non-interacting $N$-particle bosonic system, described by the $(N+1) \times (N+1)$ dimensional density matrix (\ref{Eq: density matrix}). Then, we formulated an effective density matrix  (\ref{Eq: effective density matrix}), which contains all the information about the many-particle system relevant for the calculation of one-particle observables. In this framework we derived the generelized Josephson equations (\ref{Eq: generalized Josephson1})-(\ref{Eq: generalized Josephson3}) and their general solution for the evolution of the effective density matrix elements, \changes{see Appendix~\ref{Appendix: Generalised Josephson equations}}. This allows us to describe various regimes of many-particle dynamics:
in what follows, we first use the generalized Josephson equations to investigate a bosonic system in thermal equilibrium. 
\changes{The thermalization process, which leads to this state, is not part of the present discussion.
However, we briefly elaborate on that in Appendix~\ref{Appendix: Thermalization in a bosonic gas}.}

In particular, we will utilize the description of the many-particle system by a canonical ensemble to determine the parameters of the effective density matrix in terms of the temperature $T$, the particle number $N$ and the energy difference  $\Delta \tilde E=\tilde{E}_{1} -\tilde{E}_{0}$. \changes{Afterwards, we extend the discussion to the case of non-equilibrium leading to oscillatory dynamics.} These oscillations are studied in the pure state and in the generalized cases. Moreover, the corresponding behaviour of the degree of fragmentation $1 - f$ is analyzed. This will allow us to investigate how the interplay of temperature, particle number, and the geometry of the double-well potential affects the particle dynamics.

\subsection{Effective density matrix in thermal equilibrium}
\label{subsec: Effective density matrix in thermal equilibrium}

In the following, we want to describe a closed many-particle bosonic system in thermal equilibrium of finite temperature $T$. In thermodynamics and statistical physics this is typically done by assuming a canonical ensemble with the density matrix 
\begin{equation}
    \hat{\rho} = \mathcal{Z}^{-1}e^{-\beta \hat{H}}\,,
    \label{eq: canonical rho}
\end{equation}
where we introduced $\beta = 1/k_\mathrm{B}T$, and $\hat{H}$ is given by Eq.~(\ref{eq: many-particle hamiltonian H}) \cite{landau2013statistical}. Moreover, the partition function $\mathcal{Z} = \mathrm{tr} (e^{-\beta \hat{H}})$ ensures the normalization $\mathrm{tr} \hat{\rho} = 1$. 

Projecting the density operator (\ref{eq: canonical rho}) to the basis of states $|\Psi_{N_{1}}\rangle$, as done in Eq.~(\ref{Eq: density matrix}), we can read off the density matrix elements
\begin{eqnarray*}
    p_{N_{1}\tilde{N}_{1}} &=& \langle \Psi_{N_{1}}| \mathcal{Z}^{-1} e^{-\beta \hat{H}} |\Psi_{\tilde{N}_{1}}\rangle \\
    &=& \mathcal{Z}^{-1} e^{-\beta(N-N_{1}){\tilde{E}}_{0} - \beta N_{1}{\tilde{E}}_{1}}\delta_{N_{1}, \tilde{N_{1}}} \,.
\end{eqnarray*}
We find that the original density matrix $\hat{\rho}$ is diagonal, which also results in a diagonal effective density matrix $\hat{\rho}_\mathrm{e}$ in the $|\psi_{0/1} \rangle$ basis with 
\begin{eqnarray}
   \alpha_{00} &=& \mathcal{Z}^{-1}\sum_{N_{1} = 0}^{N}(N - N_{1})e^{-\beta {\tilde{E}}_{1}N_{1} - \beta  {\tilde{E}}_{0} (N-N_{1})} \nonumber\\
    &=& \frac{1 + Ne^{(N+1)\beta \Delta  {\tilde{E}}} - (N+1)e^{N\beta \Delta  {\tilde{E}}}}{(1 - e^{-\beta \Delta  {\tilde{E}}})(e^{(N+1)\beta \Delta  {\tilde{E}}} - 1)} \label{alpha00 thermal}\\
    \alpha_{11} &=& \mathcal{Z}^{-1}\sum_{N_{1} = 0}^{N}N_{1}e^{-\beta  {\tilde{E}}_{1}N_{1} - \beta  {\tilde{E}}_{0} (N-N_{1})} \nonumber\\
    &=& \frac{e^{(N+1)\beta \Delta  {\tilde{E}}} + N - (N+1)e^{\beta \Delta  {\tilde{E}}}}{( e^{\beta \Delta  {\tilde{E}}} - 1)(e^{(N+1)\beta \Delta  {\tilde{E}}} - 1)} \label{alpha11 thermal}\\
    \alpha_{01} &=& \alpha_{10} = 0, \label{alpha01 thermal}
\end{eqnarray}
where $\Delta  {\tilde{E}} = \changes{\sqrt{\Delta E^2 + V_0^2}}$\changes{, cf. Eq.~(\ref{eq: tilde E0/1}),} and consistently $\alpha_{00} + \alpha_{11} = N$. This gives rise to the effective density matrix 
\begin{equation}
       \hat{\rho}_\mathrm{e} = 
        \begin{bmatrix} 
	\alpha_{00} & 	0  \\
         0 &   \alpha_{11} \\ 
	\end{bmatrix}
 \label{eq: thermal effective matrix}
 \end{equation}
as the starting point for our investigation of the generalized Josephson effect in thermal equilibrium.

\subsection{Static solution in thermal equilibrium}
\label{Subsec: Generalized Josephson effect in thermal equilibrium}

\changes{
In this work we assume a Bose gas with negligibly small interactions.
In such a system the 
establishment of thermal equilibrium 
takes a very long time $\Delta t_\mathrm{th} \gg \hbar/\Delta E$ that is much larger than the typical time scale of dynamics in the double-well potential (see also Appendix~\ref{Appendix: Thermalization in a bosonic gas}).
In this section, we assume that the system has already undergone this thermalization process and reached thermal equilibrium.}
Then the elements of the effective density matrix $\hat{\rho}_\mathrm{e}$ are given by Eqs.~(\ref{alpha00 thermal})-(\ref{alpha01 thermal}) and we are ready to apply our formalism to describe the generalized Josephson effect in this regime. 

For that purpose we consider the effective density matrix in the left and right well basis.
Applying the transformation $\hat{T}$ from Eq.~(\ref{eq:left-right states}) to Eq.~(\ref{eq: thermal effective matrix}), we obtain
\begin{multline}\changes{
   \hat{\rho}_{\mathrm{e} LR}(V_0) = \frac{N}{2}\begin{bmatrix} 
	1 &   0  \\
    0 & 	1  \\ 
\end{bmatrix}  + \frac{\delta N_{01}}{2\sqrt{\Delta E^2+V_0^2}}
 \begin{bmatrix} 
	-V_0 &  \Delta E   \\
         \Delta E & 	 V_0  \\ 
	\end{bmatrix}, }\label{eqn:density_matrix_thermal_eq}
\end{multline}
where we introduced the population imbalance $\delta N_{01}(N, \beta \Delta {\tilde{E}})=\alpha_{00} - \alpha_{11}$ between ground and excited state.
Using Eq.~(\ref{eqn:density_matrix_thermal_eq}) as a specific initial condition for these solutions we find 
\begin{equation}
  \changes{ Z(t) = -\frac{V_0}{\Delta \tilde{E}}\frac{\delta N_{01}}{N}\,\,,\quad A(t) = \frac{\Delta E}{\Delta \tilde{E}}\frac{\delta N_{01}}{2}\,\,,\quad \theta(t)=0\,,}
   \label{eq: asym delta N imbalance equilibrium}
\end{equation}
such that no oscillations between the wells occur. 
In consequence, Eq.~(\ref{eqn:density_matrix_thermal_eq}) is not only the initial condition but a static solution for the effective density matrix, i.e., for the dynamics of the system.
\changes{
We recognize that Eqs.~(\ref{eq: asym delta N imbalance equilibrium}) represent the unique solution of the generalized Josephson equations for $N$ particles in a given double-well geometry at temperature $T$.}

\begin{figure}[b!]    \includegraphics[width=2.9 in]{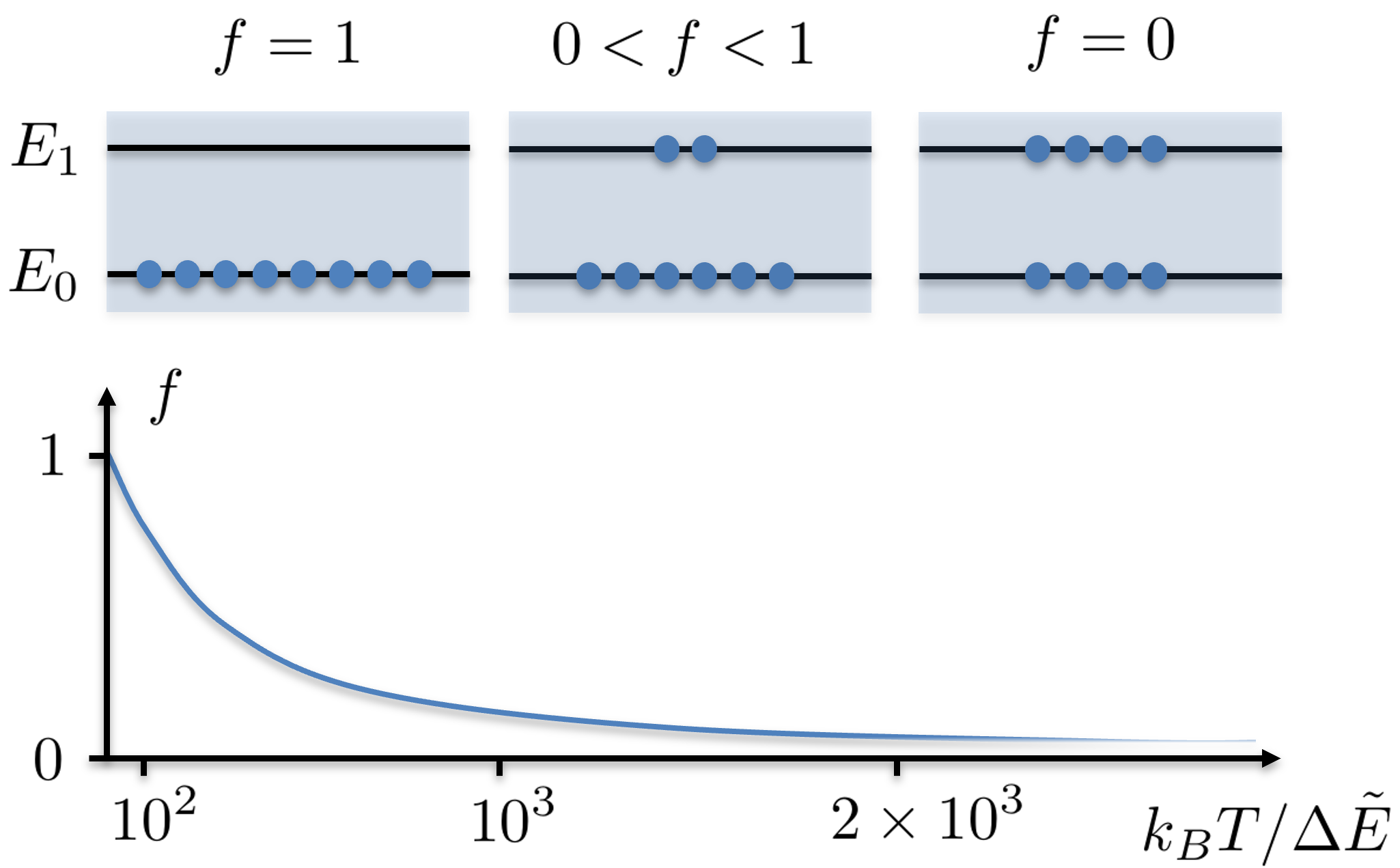}
   \caption{ Degree of condensation $f = \delta N_{01}/N$ as a function of temperature $T$ in thermal equilibrium. Above the populations of the ground ${\tilde{E}}_0$ and excited states ${\tilde{E}}_1$ are shown schematically.}
   \label{Fig: degree of condensation}
\end{figure}

From the solution (\ref{eqn:density_matrix_thermal_eq})
we find, that in thermal equilibrium, the degree of fragmentation $1-f = 1-\delta N_{01}/N$ equals the population imbalance between the excited state $|\psi_{1}\rangle$ and the ground state $|\psi_{0}\rangle$. 
In this case $f$ reaches from $f = 0$ for $k_\mathrm{B}T \gg \Delta {\tilde{E}}$ to $f = 1$ for $k_\mathrm{B}T \ll \Delta {\tilde{E}}$, as shown in Fig.~\ref{Fig: degree of condensation}. The latter particularly holds true for $T=0$, where  $\alpha_{00} = N$, such that the bosons form a global BEC in the double-well system. Hence, in thermal equilibrium $f \in [0,1]$ can be interpreted as the \emph{degree of condensation}. In the literature such a parameter is also used to introduce $T \neq 0$ effects in the Gross-Pitaevskii equation phenomenologically \cite{tienedissipative}. However, in our model it arises from first principles of statistics of a many-particle quantum system and has a wider interpretation in the non-equilibrium case.

We can summarize the findings of this section with the evident statement that in the case of thermal equilibrium the solutions of the generalized Josephson equations are static and the system exhibits no Josephson oscillations. However, the insights from this section will be of great value also for the study of the non-equilibrium case in the next section.

\subsection{Josephson effect in non-equilibrium regime}
\label{Subsec: Generalized Josephson effect with additional imbalance}
\changes{We now will discuss how Josephson oscillations appear as an out-of-equilibrium phenomenon in a particular experimental scenario \cite{PhysRevLett.118.230403}.
First a cloud of atoms is thermalized to the equilibrium configuration with some population imbalance between the wells, induced by the energy difference $E_L - E_R = V_0^{i}$. 
Then, the potential step between the wells is instantly lowered to a new value $V_0^f < V_0^i$, bringing the system out of equilibrium. 
Directly after that, the system evolves according to Eqs.~(\ref{Eq: generalized Josephson1})-(\ref{Eq: generalized Josephson3}), but  with $E_L - E_R = V_0^f$.
The initial conditions for the system are given by $\hat{\rho}_{\mathrm{e} LR}(V_0^i)$ from Eq.~(\ref{eqn:density_matrix_thermal_eq}). Plugging that into the general solutions to the generalized Josephson equations we present in Appendix~\ref{Appendix: Generalised Josephson equations}, we obtain the population imbalance}
\changes{\begin{multline}
   Z(t) =  - \frac{V_{0}^f}{\sqrt{\Delta E^2 + (V_0^i)^2}}\frac{\delta N_{01}^i}{N}  \\ - \frac{V_0^i - V_0^f}{\sqrt{\Delta E^2 + (V_0^i)^2}}\frac{\delta N_{01}^i}{N} \cos \left(\frac{\Delta E t}{\hbar}\right),
   \label{eq: asym delta N imbalance}
\end{multline}
showing an oscillatory behavior. Here, for the sake of simplicity we only consider terms to linear order in the parameter $V_0^{f}/\Delta E$, which is assumed to be small in our further discussion. 
This includes the case considered in Ref.~\cite{PhysRevLett.118.230403}, where the final double-well is symmetric \mbox{$V_0^f = 0$}.} 

\begin{figure}[b!]
    \includegraphics[width=3.1 in]{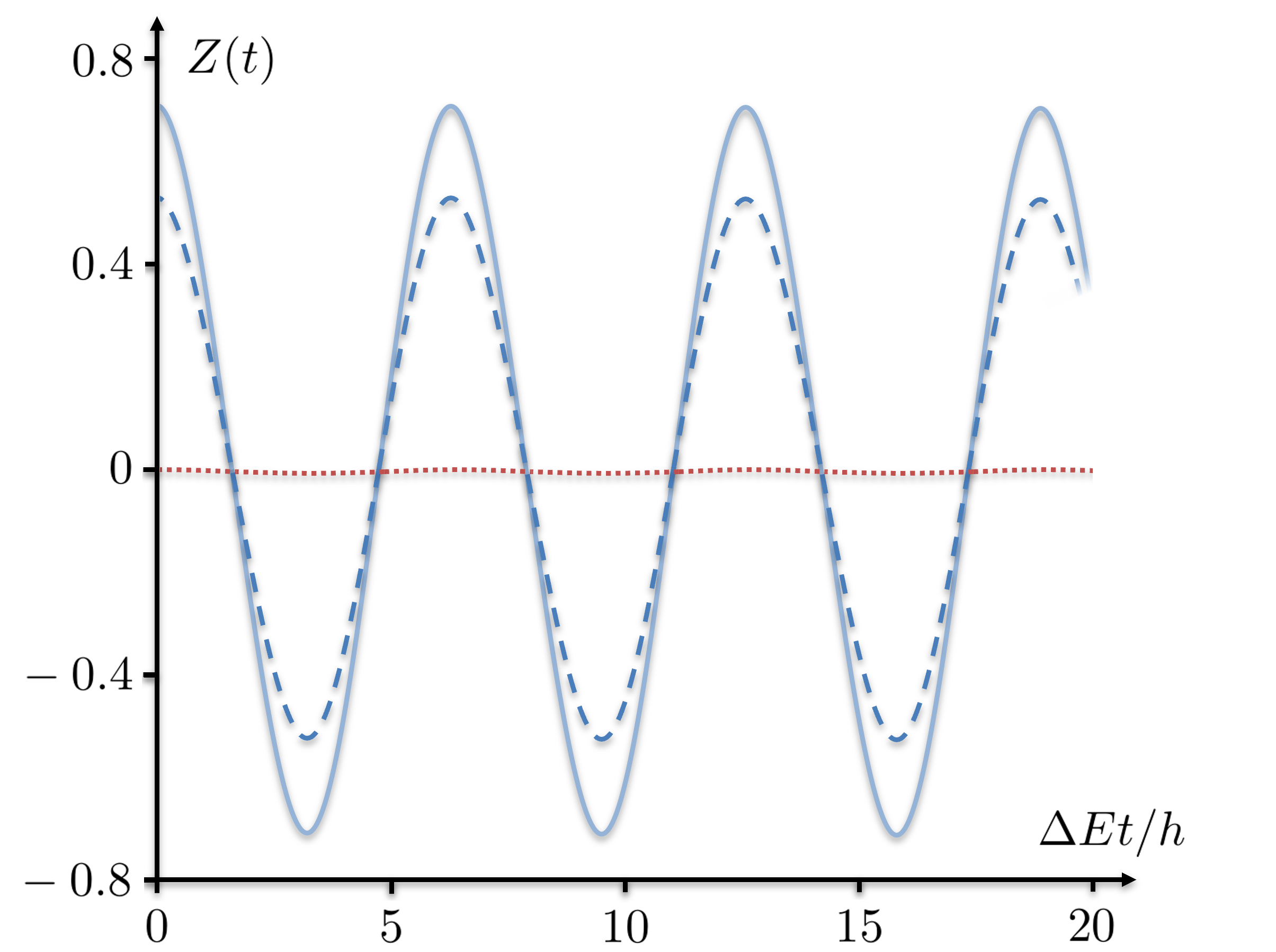}
   \caption{\changes{ Oscillations of the population imbalance $Z(t)$ as a function of dimensionless time $\Delta E t/\hbar$. The initial potential step is chosen as $V_0^i = \Delta E$ and the final potential step \mbox{$V_0^f = 0$}. The number of particles is fixed to $N=10^3$, while the temperature is  $T = 2.5 \times 10^{-4}$ K (red dotted), $T = 10^{-6}$ K (blue dashed), and $T = 10^{-8}$ K (blue solid).} }
   \label{Fig: oscillations}
\end{figure}

\changes{As can be seen in Eq.~(\ref{eq: asym delta N imbalance}), the amplitude of oscillations is proportional to the difference between the initial and final potential step $V_0^i - V_0^f$, which also quantifies how far the system is from the initial equilibrium state. 
In the limit $V_0^i \gg \Delta E$ we observe the maximum possible amplitude of oscillations $\delta N_{01}^i/N$ for given $N$, $\Delta E$ and $T$, see Fig.~\ref{Fig: oscillations} for particular temperatures. In the opposite case if $V_0^i \ll \Delta E$ the amplitude of Josephson oscillations is proportional to a small ratio $(V_0^i - V_0^f)/\Delta E$. }

\changes{The $f$ parameter in the non-equilibrium regime reads $f = \delta N_{01}^i/N$, where $\delta N_{01}^i = \alpha_{00}^i - \alpha_{11}^i$ is the initial population imbalance between the ground  and first excited state with the energies $\tilde{E}_0^i$ and $\tilde{E}_1^i$, respectively. Moreover, the $f$ parameter remains independent of the new potential step $V_0^f$ and is preserved from the initial thermal equilibrium condition, discussed in Sec.~\ref{Subsec: Generalized Josephson effect in thermal equilibrium}.}
\changes{In general the values $V_0^i$, $V_0^f$ and $\Delta E$ are known from the experimental setup \cite{PhysRevLett.118.230403}, which allows us to deduce $f$ from the oscillation amplitude. 
The $f$ parameter depends only on the number of bosons $N$ and $\Delta \tilde{E}^i/(k_B T)$.
Thus, by the measurement of the amplitude of Josephson oscillations in the Bose gas it is possible to determine the thermodynamic temperature 
it had before the potential step was lowered $V_0^i\to V_0^f$.}

\changes{
The pure state case $f=1$, for which standard Josephson equations hold, is realized for $k_B T \ll \Delta \tilde{E}^i$, allowing for the maximum possible oscillation amplitude. The opposite regime of} $f=0$ can be achieved either in the case of high temperatures or in the case of nearly degenerated \changes{initial energy levels $\tilde{E}_{0/1}^i$ in the double-well, i.e, $\Delta \tilde{E}^i \approx 0 $}. Nearly degenerated energy levels appear for very high potential barriers between the two wells. In such a geometry the populations of each well become weakly coupled, as can be seen in the Eqs.~(\ref{eq: psiLR single particle}) and (\ref{eqn: constants}).  This is exactly the regime, where the system can be seen as two \emph{separate}, weakly interacting systems (often referred to as coupled BECs), as it is done, e.g., in the Bose-Hubbard model \cite{pitaevskii2016bose, PhysRevA.55.4318}.

\changes{The equilibrium position of the oscillations (\ref{eq: asym delta N imbalance}) is given by the averages $\langle Z(t) \rangle_t = - V_{0}^f/\sqrt{\Delta E^2 + (V_0^i)^2}\times \delta N_{01}^i/N$ and $\langle \theta (t) \rangle_t = 0$
over one period of oscillations.
We see that this equilibrium position depends on the initial value of potential step $V_0^i$, implying that the system `remembers' the initial condition.
By direct comparison with the Eqs.~(\ref{eq: asym delta N imbalance equilibrium}) we find that this equilibrium position does not represent a static thermal equilibrium solution of the generalized Josephson equations for the new double-well with the potential step $V_0^f$.
Thus, to reach the new static equilibrium the system must `forget' its initial condition during the process of thermalization. 
Therefore, this process must not only cause a damping of the oscillations but also has to
shift the equilibrium position to  the position of a, yet undefined, new static thermal equilibrium. 
While this process itself
is not accessible within our model directly (see also Appendix~\ref{Appendix: Thermalization in a bosonic gas}), 
we can deduce some conclusions about its final state.
The exact nature of the thermalization process defines to which final thermal equilibrium state the system tends to. All these possible final thermal equilibrium states are labeled by their degree of fragmentation $1-f$. For each of those $f\in [0,1]$ there is a unique static solution of the generalized Josephson equations (\ref{eqs: generalized with f1}) - (\ref{eqs: generalized with f2}) and the degree of fragmentation for the final equilibrium state does not necessarily coincides with the initial one. Assuming a fixed degree of fragmentation $1-f = 1 - \delta N_{01}^i/N$, inherited from the initial condition, one would need to allow for a loss of energy. Otherwise, assuming the energy to be conserved, one has to account for an increase of the degree of fragmentation to allow for thermalization to happen. This means that the system is less coherent after an adiabatic thermalization process.}

\changes{A more detailed analysis of the thermalization mechanism in dependence of the underlying physical processes can be an interesting future perspective of our study.
}

\subsection{Implications for cold atom experiments}
\label{Subsec: Implications for cold atom experiments}

\changes{In what follows, we consider 
an $N$-particle system of initial 
temperature $T$ 
and discuss the implications of our model for different experimental setups.}
In Table~\ref{Table_exp} we give the upper limit for the \changes{amplitude of Josephson oscillations (\ref{eq: asym delta N imbalance}), determined by the initial degree of condensation $f = \delta N_{01}^i/N$} for experimentally relevant scenarios, related to typical temperatures and cooling techniques \cite{Cronin_2009}.
\changes{The corresponding oscillations for $N=10^3$ and different temperatures are visualized in Fig.~\ref{Fig: oscillations}.}

On its way to ultra-cold temperatures the many-particle bosonic system undergoes different cooling stages of characteristic temperatures.
The first stage is the formation of a longitudinal supersonic beam with a narrow velocity distribution (about $T \propto 8$ K) \cite{smalley1977molecular}.  

If lower temperatures are pursued, 
the beam can be placed in a magneto-optical trap (MOT)  and can be laser-cooled to temperatures of
$T \propto 2.5 \times 10^{-4}$ K \cite{phillips1991optical}.
In this regime the  $f$ parameter sensitively depends on the particle number, ranging from $N \propto 10^{3} - 10^{6}$ in typical experiments \cite{Smerzi1997, rudolph2015high, sewell2010atom, Salasnich_1999, Albiez_2005, PhysRevA.55.4318,PhysRevLett.118.230403}.
\changes{While for $10^3$ particles we see that the amplitude of Josephson oscillations is restricted to maximally $0.5 \%$ of all particles, for $10^6$ this amplitude is barely restricted.}

At the next cooling stage the atoms are placed in an optical dipole trap reaching  typical temperatures in the range of $T \propto 10^{-6} $ K \cite{prentiss1989effect}. 
\changes{For low particle numbers $\propto 10^3$, in this regime, it is necessary to use not the standard, but the generalized Josephson equations with a degree of fragmentation of $1 - f \approx 0.25$ (see Table~\ref{Table_exp}). To describe such a cold (but not yet ultra-cold) Bose gas correctly is of interest for modern precision metrology \cite{unnikrishnan2018bose}. 
}

Finally, evaporative cooling and adiabatic expansion can lead to the BEC formation at $T \propto 10^{-8}$ K \cite{anderson1995observation} with almost all particles  condensed in the ground state, such that $\delta N^i_{01} \approx N$.
In this regime the system is almost fully coherent and can be represented by a single wave-function to good approximation. Here, the standard Josephson equations are sufficient to describe the dynamics of the many-particle system.
\changes{In the BEC regime our model puts no significant restriction on the amplitude of Josephson oscillations, in agreement with the experimental observations \cite{Smerzi1997, sewell2010atom, Albiez_2005, PhysRevA.55.4318, PhysRevLett.118.230403}.} 

\begin{table}[t!]
\caption{\label{tab:table3} 
Limit on maximum \changes{amplitude of Josephson oscillations $\delta N_{01}^i/N$}  for typical temperatures $T$ and particle numbers $N$, related to specific experimental scenarios. An exemplary oscillation frequency of  \changes{$\Delta E/\hbar=10^3\,\mathrm{rad/s}$} out of the typical range of \changes{$10^2\, \mathrm{rad/s} - 10^4\, \mathrm{rad/s}$} \cite{Smerzi1997, sewell2010atom, Albiez_2005, PhysRevA.55.4318,PhysRevLett.118.230403}.}
\changes{\begin{tabular}{lll}
\noalign{\smallskip}\hline\noalign{\smallskip}
Temperature regime  \enspace  &  $N$\enspace \enspace \enspace & \changes{$f = \delta N_{01}^i/N$}  \\
\noalign{\smallskip}\hline\noalign{\smallskip}
 Optical molasses & $10^3$ & 0.0051 \\
     or MOT        & $10^4$ & 0.0508 \\
  $T = 2.5 \times 10^{-4}$ K & $10^5$ & 0.4443 \\
 & $10^6$ & 0.9345\\
\noalign{\smallskip}\hline\noalign{\smallskip}
   Collimated beam        & $10^3$ & 0.7401 \\
    (transverse) & $10^4$ &  0.9739\\
    $T = 10^{-6}$ K & $10^5$ & 0.9974\\
   & $10^6$ & 0.9997\\
\noalign{\smallskip}\hline\noalign{\smallskip}
     & $10^3$ & 0.9983 \\
BEC $T = 10^{-8}$ K & $10^4$ & 0.9998 \\
  \enspace  & $10^5$ & 0.9999 \\
 & $10^6$ & 1.0000 \\
 \noalign{\smallskip}\hline\noalign{\smallskip}
\end{tabular}}
\label{Table_exp}
\end{table}

\section{Conclusions}
\label{sec: Conclusions}

In this paper we derived a generalization of the standard Josephson equations, which can be used to describe bosonic many-particle systems in the non-coherent regime, apart from Bose-Einstein condensation. In particular, we apply this formalism to study a system of $N$ quantum particles in an asymmetric double-well potential at finite temperatures. For this purpose we first construct an effective density matrix, which allows us to calculate the expectation values of one-particle operators. For this density matrix, we derive the generalized Josephson equations as the central part of our theory. These equations define the evolution of the population imbalance, the phase difference and a newly introduced mixing parameter, which together provide a full description of the bosonic system on the level of one-particle operators. The newly introduced mixing parameter allows us to investigate the many-particle bosonic system, that not necessarily has to be in the BEC phase.

The generalization of Josephson equations leads to additional physical effects. 
To analyze these effects we introduce an additional parameter $f\leq 1$ 
with $1-f$ having a meaning of the degree of fragmentation. The ultimate case $f=1$
corresponds to the pure state of the system described by the standard Josephson equations, as they are derived in literature \cite{Smerzi1997, RevModPhys.71.463, Salasnich_1999, cordes2001tunnelling, PhysRevA.55.4318}. 

The approach presented in this paper does not require any restrictions on the initial many-particle density matrix, up to its defining properties, implying  $f \in [0,1]$. Hence, it is suitable to describe a wide range of physical scenarios beyond the pure state case. In thermal equilibrium of finite temperature $T > 0$ the approach yields a static solution with constant population imbalance and zero phase difference between the wells. In this regime the parameter $f=\delta N_{01}/N$  equals the fractional population imbalance between the ground and excited energy eigenstates and, hence, has the meaning of the degree of condensation.

\changes{To discuss the non-equilibrium regime we considered an initially thermalized Bose gas in an asymmetric double-well potential. Then, in the modeled experimental  scenario \cite{PhysRevLett.118.230403},  
the potential step between the wells
is instantly lowered. This leads to oscillatory dynamic in a double-well system.
We found that the oscillation amplitude depends on temperature $T$, total number of particles $N$, energy difference $\Delta E$, as well as the initial and final potential steps $V_0^i$ and $V_0^f$. 
In experiment one can access the values of $V_0^i$, $V_0^f$, $\Delta E$ and $N$, implying that the knowledge of the amplitude of the Josephson oscillations may allow to determine the temperature of the system. This opens up an intriguing possibility of quantum thermometry.}

\changes{We found that for a fixed double-well geometry the amplitude is limited by the $f$ parameter, which coincides with the initial degree of condensation $\delta N_{01}^i/N$ of the system before the potential step was lowered.
This restriction becomes recognizable for cold (but not ultra-cold) Bose gases at temperatures $T \geq 10^{-6}$ K and vanishes in the BEC regime of $T \propto 10^{-8}$ K.
This analysis highlights that the Josephson effect as a quantum interference phenomenon is more pronounced for ultra-cold Bose gases \cite{pitaevskii2016bose}.
However, the presented results complement this well-known statement by a quantitative discussion of the suppression of Josephson oscillations in the finite temperature regime.
While this conclusion fits the intuition about the Josephson effect, that is expected to vanish at higher temperatures, we want to point out, that this results sensitively depend on the experimental setup, i.e., on how non-equilibrium is obtained.
}

The matter of this paper is the generalization of the Josephson equations based on the statistical properties of the quantum system, which allows to go beyond the BEC regime. However, the applicability of our model is limited by the assumption \changes{that the interactions between the bosons are negligibly small, which leads to a long time of thermalization in comparison to oscillation time scales. Moreover, we assumed that the trapped Bose is strongly confined in all but one direction,} the total number of bosons remains constant and the system is closed, i.e., does not interact with the environment. For instance, by including interaction between the bosons, one will have an additional energy scale in the model, while implying higher-order correlation between the bosons \changes{and allowing for thermalization in a three-dimensional analysis}. These may lead to a significant modification of the model defining the possible directions in which the present study could be extended.

\acknowledgements{The authors want to thank Dorothee Tell, Yuriy M. Bidasyuk, Claus Lämmerzahl, and Andrey Surzhykov for comments and valuable discussions. Funded by the Deutsche Forschungsgemeinschaft (DFG, German Research Foundation) under Germany’s Excellence Strategy—EXC 2123 QuantumFrontiers—390837967.}

\appendix

\section{\changes{Thermalization in a bosonic gas}}
\label{Appendix: Thermalization in a bosonic gas}

\changes{In this Appendix we briefly address the constraints on our model 
under which the Bose gas can be assumed to be an effective 1D system of non-interacting particles.
We address how such a system can reach thermal equilibrium and state the 
conditions under which thermalization processes can be neglected on the time scales of Josephson oscillations.}

\changes{The effective 1D model, considered in this paper, holds for Bose gases which are strongly confined in the other two directions \cite{mazets2010thermalization}. This confinement translates into large energies $\hbar\omega_{z} = \hbar\omega_{y} = \hbar\omega_\perp\gg \Delta E$, needed to excite a perpendicular state. In order to suppress this excitations in a system of finite temperature $T$, the inequality $\hbar \omega_\perp > k_B T$ must hold.
In this case, the wavefunction can be decoupled into a product of a transverse ground state and a 1D state $\psi(x, t)$ we are interested in.
}

\changes{For establishment of thermal equilibrium one would need to have a redistribution of energy leading to thermalization. 
However, in a 1D collision of two identical bosons no energy exchange and therefore no thermalization would happen.
This redistribution can be achieved only by including 3D inter-particle collisions, which would populate the transverse excited states, and which are neglected in the Hamiltonian~(\ref{eq: many-particle hamiltonian H}). 
These collisional interactions can be present but negligible, as long as the linear particle density $n_{1D}$ 
and the scattering length $a_s$ of the bosons satisfy $n_{1D}a_s \ll 1$ \cite{mazets2010thermalization}.
In this case the time scale of thermalization $\Delta t_{th}$ is much larger than the period $h/\Delta E$ of Josephson oscillations. This is in agreement with 
 the analysis in Sec.~\ref{sec:many paricle description}
and Sec.~\ref{sec: Generalized Josephson effect}. In particular, within this approximation, we neglect damping due to reestablishment of thermal equilibrium in Sec.~\ref{Subsec: Generalized Josephson effect with additional imbalance}.}

\changes{For a detailed study on the thermalization process, see Ref.~\cite{mazets2010thermalization}.}

\section{Standard Josephson solutions}
\label{Appendix: Standard Josephson equations}

In this Appendix the technical details of solving the pure state Josephson equations~(\ref{Eq: pure state Josephson1}),(\ref{Eq: pure state Josephson2}) are discussed. To simplify the problem, we divide these equations following the rule $\frac{dZ}{dt}\left(\frac{d\theta}{dt}\right)^{-1}=\frac{dZ}{dt}\frac{dt}{d\theta}=\frac{dZ}{d\theta}$
 to obtain a differential equation for the $Z(\theta)$ function. Then we introduce new functions $x = \cos \theta$ and $Z = \sin y$, in terms of which the differential equation reads
\begin{equation*}
   \frac{dx}{dy} = x\tan y - \frac{E_{L} - E_{R}}{2K}.
\end{equation*}
In the following we will denote the dimensionless ratio in this equation as $\delta = (E_{L} - E_{R})/(2K)$. In the case of the asymmetric double-well (\ref{eqn: constants}) we find $\delta = - V_0/\Delta E$.
The general solution is $x(y) = \frac{\sqrt{1 - \Delta \rho_\mathrm{s}^{2}}}{\cos y} + \frac{c_{2}\delta}{\cos y} - \delta \tan y$, so $\cos \theta = \frac{\sqrt{1- \Delta \rho_\mathrm{s}^{2}} + c_{2}\delta}{\sqrt{1 - Z(t)^{2}}}  - \delta \frac{Z}{\sqrt{1-Z^{2}}}$. Substituting this result into the first equation~(\ref{Eq: pure state Josephson2}), one gets:
\begin{equation*}
   \frac{dZ}{\sqrt{1 + \beta Z- \gamma Z^{2}}} = \alpha dt \,,
\end{equation*}
where the following notations are introduced $\beta = \frac{2\delta\left(\sqrt{1 - \Delta \rho_\mathrm{s}^{2}} + c_{2}\delta \right)}{\Delta \rho_\mathrm{s}^{2} -2c_{2}\delta \sqrt{1-\Delta \rho_\mathrm{s}^{2}}- c_{2}^{2}\delta^{2}}$, $\gamma = \frac{1 + \delta^{2}}{\Delta \rho_\mathrm{s}^{2} -2c_{2}\delta \sqrt{1-\Delta \rho_\mathrm{s}^{2}}- c_{2}^{2}\delta^{2}}$ and $\alpha = \frac{2K}{\hbar}\sqrt{\Delta \rho_\mathrm{s}^{2} -2c_{2}\delta \sqrt{1-\Delta \rho_\mathrm{s}^{2}}- c_{2}^{2}\delta^{2}}$. Performing the shift $\tilde{Z} = Z - \frac{\beta}{2\gamma}$ one obtains the solution for $Z(t)$ and $\theta(t)$:
\begin{widetext}
\begin{align*}
   & Z(t)   =
    \frac{\delta(c_{2}\delta + \sqrt{1 - \Delta \rho_\mathrm{s}^{2}})}{1 + \delta^{2}} +  \frac{\sqrt{\Delta \rho_\mathrm{s}^{2} + \delta(\delta - c_{2}^{2}\delta - 2c_{2}\sqrt{1 - \Delta \rho_\mathrm{s}^{2}}))}}{1 + \delta^{2}} \sin\left(\frac{2K^{\mathrm{(a)}}}{\hbar}\sqrt{1 + \delta^{2}} t + \phi_{0\mathrm{s}} + \delta \alpha \right) ,\\
& \theta(t)  = \arccos \left(\frac{\sqrt{1- \Delta \rho_\mathrm{s}^{2}} + c_{2}\delta}{\sqrt{1 - Z(t)^{2}}}  - \delta \frac{Z(t)}{\sqrt{1-Z(t)^{2}}}\right).
\end{align*}
 \end{widetext}
To obtain the solution in the case of a symmetric double-well, one needs to set $\delta = 0$ in the expressions above. The latter solution contains the two integration constants $\Delta \rho_\mathrm{s}$ and $\phi_{0\mathrm{s}}$, while in the asymmetric case the two additional constants $c_{2}$ and $\delta \alpha$ appear.

To linear order in $\delta= - V_0/\Delta E$, 
i.e, for a small asymmetry of the double-well potential, we have
\begin{multline*}
    Z(t) =
   - \frac{V_0}{\Delta E} \sqrt{1 - \Delta \rho_\mathrm{s}^{2}} \\+ \left(\Delta \rho_\mathrm{s} + c_{2}\frac{V_0}{\Delta E}\frac{\sqrt{1 - \Delta \rho_\mathrm{s}^{2}}}{\Delta \rho_\mathrm{s}} \right) \sin\psi_\mathrm{t} \\- \frac{V_0}{\Delta E} \alpha \Delta \rho_\mathrm{s} \cos \psi_\mathrm{t},
\end{multline*}
\begin{multline*}
\theta(t)  = \arccos \left(\frac{\sqrt{1 - \Delta \rho_\mathrm{s}^{2}} }{\sqrt{1-\Delta \rho_\mathrm{s}^{2}\sin^{2}\psi_\mathrm{t}}}\right) \\ - \frac{V_0}{\Delta E} \frac{\Delta \rho_\mathrm{s}^{3}\sin \psi_\mathrm{t} \cos \psi_\mathrm{t} - c_{2}\cos \psi_\mathrm{t} - \alpha \Delta \rho_\mathrm{s}^{2}\sqrt{1 - \Delta \rho_\mathrm{s}^{2}}\sin \psi_\mathrm{t} }{\Delta \rho_\mathrm{s}(1 - \Delta \rho_\mathrm{s}^{2}\sin^{2}\psi_\mathrm{t})},
\end{multline*}
where we introduced the notation $\psi_\mathrm{t} = \phi_{0\mathrm{s}} - \Delta E t/\hbar$.

\section{Interpretation of effective density matrix}
\label{Appendix: Interpretation of effective density matrix}

In this Appendix we will establish the interpretation of the effective density matrix $\hat{\rho}_{e}$, which determines expectation values of one-particle operators, cf. Eq.~(\ref{Eq: rho e definition}). Such operators project the Hilbert space of the many-particle states $|\Psi_{N_{1}}\rangle$ to a two-dimensional space of single-particle states $|\psi_{0/1}\rangle$. Now we will prove that such projection corresponds to reducing the original density matrix $\hat{\rho} = \sum_{N_{1}\tilde{N}_{1}}p_{N_{1}, \tilde{N}_{1}}|\Psi_{N_{1}}\rangle \langle \Psi_{\tilde{N}_{1}}|$ to a reduced single-particle density matrix $\hat{\rho}_{1}$, as used, e.g., in Ref.~\cite{erdahl2012density}. 

The system, which is the whole $N$-particle ensemble, is now divided into two subsystems $\hat{H} = \hat{H}_{1} \oplus \mathbb{1} + \mathbb{1} \oplus \hat{H}_{B} $, which are a single 1st particle and a bath of all other $N-1$ particles. The whole ensemble is described by the $|\Psi_{N_{1}}\rangle$ states from Eq.~(\ref{whole_state}), while the energy eigenbasis of the bath reads 
\begin{multline*}
|\Phi_{n_{1}}\rangle = \frac{1}{\sqrt{(N - 1)!(N - 1 - n_{1})!n_{1}!}} \int dx_{2}...dx_{N} \times \\ \times \sum_{j(n_{1})}  \psi_{\sigma_{j}^{n_{1}}(2)}(x_{2})...\psi_{\sigma_{j}^{n_{1}}(N)}(x_{N})\changes{|x_{2},\dots,x_{N}\rangle}, 
\end{multline*}
in a complete analogy with the notations in Eq.~(\ref{whole_state}).

In order to extract the important information about the state of a 1st particle, one needs to trace the total density matrix $\hat{\rho}$ over the bath states $|\Phi_{n_{1}}\rangle$
\begin{multline*}
\hat{\rho}_{1} = \mathrm{tr}_{\mathrm{B}}\hat{\rho} \\ = \sum_{n_{1} = 0}^{N - 1}\sum_{N_{1} = 0}^{N}\sum_{\tilde{N}_{1} = 0}^{N} p_{N_{1} \tilde{N}_{1}}(t) \langle \Phi_{n_{1}}|\Psi_{N_{1}}\rangle_{\mathrm{B}} \langle \Psi_{\tilde{N}_{1}}| \Phi_{n_{1}}\rangle_{\mathrm{B}},
\end{multline*}
where $\langle \Phi_{n_{1}}|\Psi_{N_{1}}\rangle_{\mathrm{B}}$ is a scalar product in the bath Hilbert space only. Hence, it is an element of the single-particle Hilbert space in the $|\psi_{0/1}\rangle$ basis. 
Due to the orthogonality of the one-particle states, 
the scalar product $\langle \Phi_{n_{1}}|\Psi_{N_{1}}\rangle_{\mathrm{B}}$ does not vanish only for values of $n_{1} = N_{1} - 1,N_{1}$. The scalar product $\langle \Psi_{\tilde{N}_{1}}| \Phi_{n_{1}}\rangle_{\mathrm{B}}$ is non-zero only for $n_{1} = \tilde{N}_{1} - 1,\tilde{N}_{1}$.
Then the expression above can be simplified to just one summation over $N_{1}$ with four terms, which can be calculated by using the explicit expressions of $|\Psi_{N_{1}}\rangle$ and $|\Phi_{n_{1}}\rangle$. For instance, the product with $n_{1} = N_{1}$ reads
\begin{multline*}
\langle \Phi_{N_{1}}| \Psi_{N_{1}}\rangle_{\mathrm{B}} = \frac{1}{\sqrt{N(N - N_{1})}}\sum_{i=1}^{N-N_{1}}|\psi_{0}(x_{i})\rangle \\ = \sqrt{\frac{N-N_{1}}{N}}|\psi_{0}(x)\rangle\,.
\end{multline*}

The last equality in the expression above holds only effectively, i.e., for the calculation of one-particle observables $\hat{O}$, introduced in Sec.~\ref{subsec:effective density matrix description}. Without referring to the particular type of observables, this approximation can be treated as the mean-field approach with one averaged coordinate $x$ instead of many-particle coordinates $\{x_{i}\}$. The same can be done for the other term
\begin{equation*}
\langle \Phi_{N_{1} - 1}| \Psi_{N_{1}}\rangle = \frac{1}{\sqrt{NN_{1}}}\sum_{i=1}^{N_{1}}|\psi_{1}(x_{i})\rangle  = \sqrt{\frac{N_{1}}{N}}|\psi_{1}(x)\rangle.
\end{equation*}

This simplification yields the reduced single-particle density matrix
\begin{multline*}
\hat{\rho}_{1} = \frac{1}{N} \left(\alpha_{00}|\psi_{0}\rangle \langle \psi_{0}| + \alpha_{01}|\psi_{1}\rangle \langle \psi_{0}| \right. \\ \left. + \alpha_{10}|\psi_{0}\rangle \langle \psi_{1}| + \alpha_{11}|\psi_{1}\rangle \langle \psi_{1}|\right)\,,
\end{multline*}
where the $\alpha_{ij}$ coefficients are the same as defined in Sec.~\ref{subsec:effective density matrix description}. So we conclude that $\hat{\rho}_\mathrm{e} = N\hat{\rho}_{1}$.

\section{Degree of fragmentation}
\label{Appendix: Degree of fragmentation}

In the Appendix we will show how to obtain the generalized Josephson equations in the form
\begin{eqnarray}
   \hbar \dot{Z} &=& 2K \sqrt{f^{2}-Z^{2}}\sin \theta ,
   \label{E1}\\
   \hbar \dot{\theta} &=& E_{L} - E_{R} - \frac{2KZ \cos \theta}{\sqrt{f^{2}-Z^{2}}},
   \label{E2}
\end{eqnarray}
analogous to the standard Josephson equations except of $f \neq 1$. The latter equations look simpler, than their original form with the mixing parameter $A(t)$ (\ref{Eq: generalized Josephson1})-(\ref{Eq: generalized Josephson3}), however this way the meaning of the $f$ parameter is hidden. To define $f$ one would need to trace back to the original generalized Josephson equations and apply the initial conditions for the effective density matrix $Z(t=0)$, $\theta(t=0)$ and $A(t=0)$. Now we will show how the $f$ parameter can be expressed via the integration constants, introduced in the Appendix above, while in the general case it reads $f^2 = Z(t)^2 + (2A(t)/N)^2$ and represents the pure state condition if $f=1$.

As shown in Appendix~\ref{Appendix: Generalised Josephson equations}, the general solution  for the symmetric double-well reads $Z(t) = \Delta \rho_\mathrm{s} \sin \left[\Delta E t/\hbar + \phi_{0\mathrm{s}}\right]$ and $\theta(t) = -\arctan \left[\frac{Ne^{-B_\mathrm{s}} \Delta \rho_\mathrm{s}}{2} \cos \left(\Delta E t/\hbar + \phi_{0\mathrm{s}}\right)\right]$ which leads to 
\begin{equation*}
    \cos \theta = \left[1 + \left(\frac{Ne^{-B_\mathrm{s}}}{2}\right)^{2}\left(\Delta \rho_\mathrm{s}^{2} - Z^{2}\right)\right]^{-1/2}\,.
\end{equation*}
We substitute this expression in $A(t) = e^{B_\mathrm{s}}/\cos \theta (t)$ and then plug it in the first two Josephson equations (\ref{Eq: generalized Josephson1}),( \ref{Eq: generalized Josephson2}), which yields
\begin{eqnarray*}
   \hbar \dot{Z} &=& -\Delta E \sqrt{(f^{\mathrm{(s)}})^2- Z^{2}} \sin \theta  \\
   \hbar \dot{\theta} &=& \Delta E \frac{Z \cos \theta}{\sqrt{(f^\mathrm{(s)})^2-Z^{2}}},
\end{eqnarray*}
where $(f^\mathrm{(s)})^2 = \left(2e^{B_\mathrm{s}}/N\right)^{2} + \Delta \rho_\mathrm{s}^{2}$. This immediately recovers pure state condition by taking $f^\mathrm{(s)} = 1$.

In slightly asymmetric double-well we make an educated guess that the equations~(\ref{Eq: generalized Josephson1}) - (\ref{Eq: generalized Josephson3}) can be written in a form (\ref{E1}), (\ref{E2}). Thus,  Eq.~(\ref{E1}) yields the condition
\begin{equation*}
     (f^\mathrm{(a)})^2 = Z^{2} + \left(\frac{\hbar}{\Delta E}\right)^{2}\frac{\dot{Z}^{2}}{\sin^{2}\theta} = const. 
\end{equation*}
Substituting the general solution~(\ref{D1}), (\ref{D3}) into the expression above, we find that indeed $f^{\mathrm{(a)}} = const$ and equals
\begin{multline*}
     (f^\mathrm{(a)})^2 = (f^\mathrm{(s)})^2 + \frac{V_0}{\Delta E}\frac{e^{-B_\mathrm{s}}}{N} \left[2c_{1}e^{2B_\mathrm{s}}\Delta \rho_\mathrm{s}\sin \delta \phi_0 \right. \\ \left. + N(f^{\mathrm{(s)}})^2\left(\delta B e^{B_\mathrm{s}} - N\Delta\rho_\mathrm{s} \sin \phi_{0\mathrm{s}}\right) \right].
\end{multline*}
In the same way one can prove that the equation (\ref{E2}) holds for the solutions of the generalized Josephson equations.

The $f = \sqrt{Z(t)^2 + (2A(t)/N)^2}$ parameter also appears   
when we diagonalize the effective density matrix $\hat{\rho}_{\mathrm{e}LR}$ from Eq.~(\ref{Eq: effective density matrix}) to obtain
\begin{multline}
   \hat{\tilde{\rho}}_{\mathrm{e}} = 
  \frac{N}{2}\begin{bmatrix} 
	1 + f &  0  \\
    0 & 1-f\\ 
	\end{bmatrix} \\ = \frac{N}{2}\left[(1 + f)|y_1\rangle \langle y_1 | + (1- f)|y_2\rangle \langle y_2|\right]
 \label{eq: rho e LR diagonal}
\end{multline}
in the orthonormal basis of its eigenvectors
\begin{multline}
   |y_{1}\rangle =  \left[1 + \left(\frac{N}{2A(t)}(Z(t) - f)\right)^2\right]^{-1/2}\\ \times \left[|\psi_{L}\rangle - \frac{Ne^{-i \theta(t)}}{2A(t)}(Z(t) - f)|\psi_{R}\rangle\right], \label{eqn:y1}
\end{multline}
\begin{multline}
   |y_{2}\rangle = \left[1 + \left(\frac{N}{2A(t)}(Z(t) + f)\right)^2\right]^{-1/2} \\ \times  \left[|\psi_{L}\rangle - \frac{Ne^{-i \theta(t)}}{2A(t)}(Z(t) + f)|\psi_{R}\rangle\right], \label{eqn:y2}
\end{multline}
which are now time-dependent. In the case of thermal equilibrium, the eigenstates $|y_{1/2}\rangle$ reduce to the ground and excited states $|\psi_{0/1}\rangle$ of the double-well potential. 

We recall that in Appendix~\ref{Appendix: Interpretation of effective density matrix} it has been shown that the effective density matrix $\hat{\rho}_\mathrm{e}$ is connected with the reduced density matrix of an average boson $\hat{\rho}_{1}$ in a bath of all other $N-1$ bosons  via the relation $\hat{\rho}_\mathrm{e} = N\hat{\rho}_{1}$. Thus, the effective density matrix has to obey the physical requirements for the density matrices, in particular, it has to be positive semidefinite, i.e., all its eigenvalues have to be positive. This implies $|f| \leq 1$. The $f$ parameter, therefore, can be interpreted as the pure state fraction in the many-particle bosonic system.

\changes{Notice that the term `pure state', so far used to describe an average boson in a pure single-particle state, implies the many-particle Fock state  \cite{pitaevskii2016bose}
\begin{equation}
    |\Psi \rangle = \frac{1}{\sqrt{N! 2^N}} (\hat a^{\dagger} + e^{i\Phi}\hat b^{\dagger})^{N}| vac \rangle
    \label{eq: Fock App}
\end{equation}
with $\hat a^{\dagger}$ and $\hat b^{\dagger}$ being the creation operators relative to the single-particle states $\psi_{0/1}$, acting on vacuum $| vac \rangle$. In our framework this state is characterized by a degree of fragmentation $1-f = 0$.  However, it was shown \cite{pitaevskii2001thermal, PhysRevLett.96.130404} that the state~(\ref{eq: Fock App}) can allow for a nonzero degree of fragmentation due to quantum fluctuations in strongly interacting BECs. A discussion of this  
regime requires to determine the degree of fragmentation differently, and is not the subject of our study. In particular, the definition used in \cite{pitaevskii2001thermal, PhysRevLett.96.130404} should not be confused with the one, presented in the paper.}

\section{Generalised Josephson solutions}
\label{Appendix: Generalised Josephson equations}

Here, we discuss how to obtain the solutions of the generalised Josephson equations~(\ref{Eq: generalized Josephson1})-(\ref{Eq: generalized Josephson3}). 

First, we present the key steps towards the solution for a symmetric double-well. In this case Eq.~(\ref{Eq: generalized Josephson3}) can be integrated directly and yields $A(t) = e^{B_\mathrm{s}}/\cos \theta(t)$, where $B_\mathrm{s}$ is an integration constant. Substituting this result into the other two equations, we obtain:
\begin{eqnarray*}
    \frac{2e^{B_\mathrm{s}}\Delta E}{N} \tan \theta(t) &=& -\hbar \dot{Z}(t)\\
    \frac{e^{-B_\mathrm{s}}\Delta E N}{2} Z(t) \cos^{2} \theta(t) &=& \hbar \dot{\theta}(t)\,.
\end{eqnarray*}
Taking the time derivative of the first equation and substituting $\dot{\theta}(t)$ from the second equation, one obtains an oscillatory equation for the population imbalance $Z(t)$:
\begin{equation*}
    \ddot{Z}(t) + \left(\frac{\Delta E}{\hbar}\right)^{2} Z(t) = 0 .
\end{equation*}
Its general solution is $Z(t) = \Delta \rho_\mathrm{s} \sin \left[\Delta E t/\hbar + \phi_{0\mathrm{s}}\right]$ with the integration constants $\Delta \rho_\mathrm{s}$ and $\phi_{0\mathrm{s}}$. This gives $\theta(t) = -\arctan \left[\frac{Ne^{-B_\mathrm{s}} \Delta \rho_\mathrm{s}}{2} \cos \left(\Delta E t/\hbar + \phi_{0\mathrm{s}}\right)\right]$, which with $A(t) = e^{B_\mathrm{s}}/\cos \theta(t)$ defines the most general solution of the system of the differential equations~(\ref{Eq: generalized Josephson1})-(\ref{Eq: generalized Josephson3}) with three integration constants $B_\mathrm{s}$, $\Delta \rho_\mathrm{s}$ and $\phi_{0\mathrm{s}}$.

Second, we discuss the case of an asymmetric double-well potential. The integration of Eq.~(\ref{Eq: generalized Josephson3}) in this case gives $A(t) = \frac{1}{\cos \theta(t)}e^{B + \frac{E_{R} - E_{L}}{\hbar}\int_{0}^{t}dt' \tan \theta(t')}$, where $B$ is an integration constant. Substituting this result into the other two equations, we obtain:
\begin{equation*}
    \hbar \dot{Z}(t) =  \frac{4K}{N} \tan \theta(t)e^{B + \frac{E_{R} - E_{L}}{\hbar}\int_{0}^{t}dt' \tan \theta(t')},
\end{equation*}
\begin{multline*}
  K N Z(t)  e^{-B - \frac{E_{R} - E_{L}}{\hbar}\int_{0}^{t}dt' \tan \theta(t')} \cos^{2} \theta(t) =\\ -\hbar \dot{\theta}(t) - (E_{R} - E_{L})\,.
\end{multline*}

Taking the time derivative of the first equation and substituting $\dot{\theta}(t)$ from the second equation, one obtains an oscillatory equation for the population imbalance $Z(t)$
\begin{multline*}
    \ddot{Z}(t) + \left(\frac{2K}{\hbar}\right)^{2} Z(t) \\= - \frac{4K (E_{R} - E_{L})}{N\hbar^{2}}e^{B + \frac{E_{R} - E_{L}}{\hbar}\int_{0}^{t}dt' \tan \theta(t')}\,.
\end{multline*}

We see that the Josephson oscillations are governed by two frequencies $2K/\hbar = - \Delta E \sin 2\xi/\hbar$ and $(E_{R} - E_{L})/\hbar = \Delta E \cos 2\xi /\hbar$ (see Sec.~\ref{subsec:single paricle states asym}). The $\theta(t)$ function can be found by taking time derivative of the second equation and equating the $\dot{Z}(t)$ from the two equations.  This way we obtain
\begin{multline*}
   \ddot{\theta} + \left(\dot{\theta} + \frac{E_{R} - E_{L}}{\hbar}\right)\left(2\dot{\theta} + \frac{E_{R}- E_{L}}{\hbar}\right)\tan \theta \\ + \frac{2K^{2}}{\hbar^{2}}\sin 2\theta = 0\,.
\end{multline*}
One can simplify this equation by change of variables $y(t) = \tan \theta (t)$. This gives a nonlinear equation:
\begin{multline*}
   \ddot{y} + \left(\frac{2K}{\hbar}\right)^{2}y + 3
\frac{E_{R} - E_{L}}{\hbar}y\dot{y} \\ + \left(\frac{E_{R} - E_{L}}{\hbar}\right)^{2}(1+y^{2})y= 0.
\end{multline*}

Let us look at the regime of small $E_{L} - E_{R}$, so the double-well is close to a symmetric one. Then we have $2K \approx -\Delta E$, $E_{R} - E_{L} = - V_0$ (for the latter equality see Sec.~\ref{subsec:single paricle states asym}) and $y \approx y_\mathrm{sym} + u$, where $y_\mathrm{sym}$ is a solution in symmetric case.  To linear order in $u$ and $V_0/\Delta E$ we obtain
\begin{multline*}
   \ddot{y}_\mathrm{sym} + \ddot{u}+ \left(\frac{\Delta E}{\hbar}\right)^{2}y_\mathrm{sym} + \left(\frac{\Delta E}{\hbar}\right)^{2}u \\ - 
\frac{3V_0}{\hbar} y_{sym}\dot{y}_\mathrm{sym}   = 0.
\end{multline*}

In the zeroth order of $V_0/\Delta E$  and small deviation $u$ one gets the same oscillatory solution $y_\mathrm{sym}(t)$ as in the symmetric case, discussed above. In the first order of $V_0/\Delta E$  and small deviation $u$ we obtain
\begin{multline*}
   \ddot{u} + \left(\frac{\Delta E}{\hbar}\right)^{2}u \\ = 
\frac{3 V_0}{2 \Delta E} \left(\frac{N e^{-B_\mathrm{s}} \Delta \rho_\mathrm{s}}{2} \frac{\Delta E}{\hbar} \right)^{2} \sin \left[-2\frac{\Delta E t}{\hbar} + 2\phi_{0\mathrm{s}}\right].
\end{multline*}
The solution of the last equation is
\begin{multline*}
   u(t) = - \left(\frac{N e^{-B_\mathrm{s}} \Delta \rho_\mathrm{s}}{2}\right)^{2} \frac{V_0}{2\Delta E} \sin \left[-\frac{2\Delta E t}{\hbar} + 2\phi_{0\mathrm{s}}\right] \\+ c_{1}\frac{V_0}{2\Delta E} \sin \left[-\frac{\Delta E t}{\hbar} + \phi_{0\mathrm{a}}\right]
\end{multline*}
and for $\theta(t)$ one gets
\begin{eqnarray}
       \theta(t)  &=& \arctan \left(\frac{Ne^{-B_\mathrm{s}}\Delta \rho_\mathrm{s}}{2} \cos \psi_\mathrm{t}\right) \label{D1} \\ & &+ \frac{1}{\zeta^{2}(t)}\frac{V_0}{2\Delta E}\,c_{1}\sin \left[\psi_\mathrm{t} + \delta \phi_{0}\right] \nonumber\\
    & &- \frac{1}{\zeta^{2}(t)}\frac{V_0}{2\Delta E}\left(\frac{N e^{-B_\mathrm{s}} \Delta \rho_\mathrm{s}}{2}\right)^{2} \sin \left[2\psi_\mathrm{t}\right] . \nonumber
\end{eqnarray}

This solution depends on the integration constants $B_{s}$, $\Delta \rho_{s}$ and $\phi_{0s}$ of the symmetric double-well problem (our zeroth order in $V_0/\Delta E$ solution) and on the two new integration constants $\phi_{0a}$ and $c_{1}$. Taking into account that the other integration constant $B$ should only slightly differ from the analogous constant of the symmetric double-well problem $B = B_\mathrm{s} + V_0/(2\Delta E) \delta B$ we obtain the solution for $A(t)$
\begin{multline}
   A(t) = e^{B_\mathrm{s}}\zeta(t) + \frac{V_0}{2\Delta E}N\Delta \rho_\mathrm{s} \zeta(t) \left(\sin \psi_\mathrm{t} - \sin \phi_{0\mathrm{s}}\right) \\ - \frac{V_0}{2\Delta E} \frac{e^{-2B_\mathrm{s}}}{\zeta(t)} \left(\frac{\Delta \rho_\mathrm{s} N}{2}\right)^{3}\cos \psi_\mathrm{t} \sin\left(2\psi_\mathrm{t}\right) \\ + \frac{V_0}{2\Delta E}c_{1} \frac{\Delta \rho_\mathrm{s} N}{2 \zeta(t)}\cos \psi_\mathrm{t} \sin \left(\delta \phi_{0} + \psi_\mathrm{t}\right) + \frac{V_0}{2\Delta E}e^{B_\mathrm{s}} \zeta(t) \delta B.
   \label{D2}
\end{multline}
The constants  $\phi_{0\mathrm{a}}$, $\delta B$ and $c_{1}$ can be fixed by the initial condition for the effective density matrix, which for the slightly asymmetric double-well is as well a linear function of a small parameter $V_0/\Delta E$.
 Thus, for fractional imbalance we have
\begin{multline}
  Z(t) = \left[\Delta \rho_\mathrm{s} - \frac{V_0}{2\Delta E} \frac{e^{-B_\mathrm{s}}}{N}\left((\Delta \rho_\mathrm{s} N)^{2} \sin \phi_{0\mathrm{s}} \right. \right. \\ \left. \left. - 2c_{1}e^{2B_\mathrm{s}} \sin \delta \phi_{0}-e^{B_\mathrm{s}}\delta B \Delta \rho_\mathrm{s} N \right) \right]\sin \psi_\mathrm{t} \\- 2\frac{V_0}{\Delta E} \frac{e^{B_\mathrm{s}}}{N} - \frac{V_0}{\Delta E} \frac{e^{B_\mathrm{s}}}{N}c_{1} \cos \delta \phi_{0} \cos \psi_\mathrm{t}, 
  \label{D3}
\end{multline}
where the denotation $\sqrt{1 + \left(\frac{Ne^{-B_\mathrm{s}}\Delta \rho_\mathrm{s}}{2}\right)^{2}\cos^{2}\psi_\mathrm{t}} = \zeta(t)$ was used.

The generalized solution reduces to the standard one, when the pure state condition $A(t) = N\sqrt{1 - Z(t)^{2}}/2$ is satisfied. It holds for all times $t$ when the integration constants in the case of the symmetric double-well obey the relation $1 - \Delta \rho_{\mathrm{s}}^{2} = (2/N)^{2}e^{2B_{\mathrm{s}}}$. For the asymmetric double-well, in the first order of $V_0/\Delta E$, the pure state condition for the parameters of the generalized solution reads
\begin{equation*}
     2c_{1}e^{2B_\mathrm{s}}\Delta \rho_\mathrm{s}\sin \delta \phi + Nf^{\mathrm{(s)}}\left(\delta B e^{B_\mathrm{s}} - N\Delta\rho_\mathrm{s} \sin \phi_{0\mathrm{s}}\right) = 0
\end{equation*}

In the first order of $V_0/\Delta E$  the generalized and the pure state solutions are perfectly consistent for the following choice of constants of generalized solution:
\begin{eqnarray*}
    \delta B &=& \frac{2(\Delta \rho_\mathrm{s}\sin \phi_{0\mathrm{s}} - c_{2})}{\sqrt{1 - \Delta \rho_\mathrm{s}^{2}}},\\
    c_{1} &=& \frac{2\alpha \Delta \rho_\mathrm{s}}{1 - \Delta \rho_\mathrm{s}^{2}}\sqrt{1 - \Delta \rho_\mathrm{s}^{2} + \frac{c_{2}^{2}}{\alpha^{2}\Delta \rho_\mathrm{s}^{4}}},\\
    \delta \phi_{0} &=& \arctan \left[\frac{c_{2}}{\alpha \Delta \rho_\mathrm{s}^{2} \sqrt{1 - \Delta \rho_\mathrm{s}^{2}}}\right].
\end{eqnarray*}





\bibliography{main}

\end{document}